\documentstyle[11pt,epsfig]{article}
[1999/03/29 PSNFSS v.7.2
Times font as default roman
: S Rahtz]

\newcommand \be{\begin{equation}}
\newcommand \ba{\begin{eqnarray}}
\newcommand \ee{\end{equation}}
\newcommand \ea{\end{eqnarray}}

\begin{document}

\begin{center}
{\LARGE Evidence of Intermittent Cascades\\ from
Discrete Hierarchical Dissipation in Turbulence}
\end{center}
\bigskip
\begin{center}
{\large Wei-Xing Zhou {\small$^{\mbox{\ref{igpp}}}$} and Didier
Sornette {\small$^{\mbox{\ref{igpp},\ref{ess},\ref{lpec}}}$}}
\end{center}
\bigskip
\begin{enumerate}
\item Institute of Geophysics and Planetary Physics, University of California,
Los Angeles, CA 90095\label{igpp}
\item Department of Earth and Space Sciences, University of
California, Los Angeles,
CA 90095\label{ess}
\item Laboratoire de Physique de la Mati\`ere Condens\'ee, CNRS UMR 6622 and
Universit\'e de Nice-Sophia Antipolis, 06108 Nice Cedex 2, France\label{lpec}
\end{enumerate}


\begin{abstract}

We present the results of a search of log-periodic corrections to
scaling in the moments of the energy dissipation rate in
experiments at high Reynolds number ($\approx 2500$) of
three-dimensional fully developed turbulence. A simple dynamical
representation of the Richardson-Kolmogorov cartoon of a cascade
shows that standard averaging techniques erase by their very
construction the possible existence of log-periodic corrections to
scaling associated with a discrete hierarchy. To remedy this
drawback, we introduce a novel ``canonical'' averaging that we
test extensively on synthetic examples constructed to mimick the
interplay between a weak log-periodic component and rather strong
multiplicative and phase noises. Our extensive tests confirm the
remarkable observation of statistically significant log-periodic
corrections to scaling, with a prefered scaling ratio for length
scales compatible with the value $\gamma = 2$. A strong
confirmation of this result is provided by the identification of
up to $5$ harmonics of the fundamental log-periodic undulations,
associated with up to $5$ levels of the underlying hierarchical
dynamical structure. A natural interpretation of our results is
that the Richardson-Kolmogorov mental picture of a cascade becomes
a realistic description if one allows for intermittent births and
deaths of discrete cascades at varying scales.

\end{abstract}

\section{Introduction}
\label{sec:intro}

Since Kolmogorov's presentation of his theory of three-dimensional
fully developed turbulence in 1941 (see \cite{Frisch1996} for a
modern account), there has been a vigorous continuing
investigation aimed at verifying its accuracy and testing for
possible deviations. While its main predictions ($k^{-5/3}$
spectrum of velocity fluctuations, universality of the exponents
of structure functions in the inertial range) are verified to a
good approximation, there are clear deviations. The most notable
is the nonlinear dependence of the exponents $\zeta_q$ of the
structure functions as a function of their order $q$, which is a
hallmark of multifractality \cite{ParisiFrisch}. There is also a
strong interest in testing Kolmogorov's universality assumptions
that all the small-scale statistical properties are uniquely and
universally determined by the scale $\ell$, the mean energy
dissipation rate $\bar{\epsilon}$ and the viscosity (see
\cite{Renner} for recent evidence of the contrary) and that there
is a well-defined limit at infinite Reynolds number (see
\cite{Barenblatt1995,Dubrulle1996,Sornette1998b} for alternative
scenarios).

Kolmogorov's K41 hypotheses ``were bas
ed physically on Richardson's idea
\cite{Richardson1922}
of the existence in the turbulence flow of vortices of all possible scales...''
\cite{K62} and turbulence phenomenology is often cast in terms of a
cartoon depicting a cascade of the energy introduced into the largest eddies
of size $\ell_0$
and ``cascading'' down a hierarchy of eddies of size $\ell_n=\ell_0/\gamma^n$
where $\gamma > 1$ (often taken equal to $2$ but without
particular significance: \cite{Frisch1996} bottom of page 103)
at the same rate $\epsilon$ and being eventually
removed by dissipation at the smallest scales.
To our knowledge, Novikov was
the first to point out in 1966 that, if the discrete hierarchy
of the cascade is taken seriously with a specific $\gamma$, the
structure functions
in turbulence should contain log-periodic oscillations
\cite{Novikov1966, Novikov1990}. His argument was that, if an
unstable eddy in a turbulent flow typically breaks up into smaller
eddies of about half or a third of its size ($\gamma=2$ or $3$),
but not into eddies ten or twenty times smaller, then one
can suspect the existence of a preferred scale factor $\gamma$, hence the
log-periodic oscillations. Smith et al. \cite{Smith1986}
confirmed on explicit geometrical examples involving generalizations
of the triadic
Cantor set that log-periodic corrections
to scaling arise in fractal models with lacunarity, i.e., having
a prefered scaling ratio $\gamma$ between scales.
Shell models construct explicitly a discrete scale invariant set
of equations on a hierarchy of shells $k_n= 2^n k_0$ in momentum space
whose solutions are marred by unwanted log-periodicities (see for instance
\cite{lvovimproved} for a refinement addressing the problem).

Oscillations in log-log plots of structure functions and other
quantities are often observed but do not seem to be stable and
depend on the nature of the global geometry of the flow and
recirculation \cite{Frisch1996,Anselmet1984} as well as the
analyzing procedure. In his book (\cite{Frisch1996}, pages
130-131), Frisch discusses the possible existence of such
log-periodic corrections and points out that the ``extended
self-similarity'' technique \cite{Benzi} which has significantly
improved the accuracy of exponents may be useful in particular
because it overcomes the distortions stemming from the undulations
which appear to be correlated across the different orders of
structure functions (see Appendix B).

Hints of the presence of log-periodic undulations decorating the
power laws can be observed in Fig.~8 of Ref.~\cite{Meneveau1991}
which studied the multifractal nature of turbulent energy
dissipation, in Fig.~5.1~p.58 and Fig. 8.6~ p.128 of
Ref.~\cite{Frisch1996}, Fig.~3.16~p.76 of Ref.~\cite{Arneodo1995},
Fig.~1b of Ref.~\cite{Tcheou1996} and Fig.~2b of
Ref.~\cite{Castaing1997} (to list a few). However, these barely
perceptible undulations are far from proving the existence of a
genuine discrete hierarchical cascade in three-dimensional flows,
a la Richardson-Kolmogorov, as they are in general considered as
``noise''. Furthermore, if it exists, the scale ratio $\gamma$ has
not be constrained.

A preliminary attempt has dealt with another type of turbulence,
namely freely decaying 2-D turbulence \cite{Johansen2000b}, in which
experimental evidence and
theoretical arguments suggest that the time-evolution of freely
decaying 2-D turbulence is governed by a discrete time scale
invariance rather than a continuous time scale invariance.
The number of vortices, their radius and their
separation display log-periodic oscillations as a function of time
with an average log-frequency of $\sim 4-5$ corresponding to a
preferred scaling ratio of $\sim 1.2-1.3$. Since the
statistical significance for each data set taken separately is not
good when comparing with a large variety of different
stochastic processes taken as null-hypothesis \cite{Zhou2001b},
Ref.~\cite{Johansen2000b} argued about a good
statistical significance of this log-periodic oscillations
on the basis of the coincidence of three highest peaks of
the periodogram analysis which occur at log-frequencies far from the
most probable
frequencies occurring solely from noise. An alternative possibility
is that the log-periodicity results from the
long-range correlations in fractional Gaussian
like-noise with Hurst exponent of the order of $H \approx 0.3$.
This case stresses the difficulty and complication
of determination of the significance level of a log-periodic
signal.

The theory and practice of log-periodicity and
its associated complex exponents has advanced
significantly in the last few years \cite{Sornette1998a,Lapidus2000}.
It has become clear in the
1980s that complex exponents and log-periodicity
appears as soon as a given physical problem is formulated on
a discretely hierarchical geometry or network. More interestingly,
only recently has it
been realized that discrete scale invariance (DSI) and its associated
complex exponents may appear ``spontaneously'' in Euclidean
systems \cite{Saleur1996b}, i.e., without the need for a pre-existing
geometrical hierarchy.
Systems where self-organized DSI has been reported include
Laplacian growth models \cite{Sornette1996a,Huang1997}, rupture
in heterogeneous systems \cite{Anifrani1995,critrup}, earthquakes
\cite{Sornette1995,Saleur1996a,Johansen1996,Johansen2000a,Huang2000a},
animals \cite{Saleur1996b}, biased diffusion in percolating systems
\cite{Stauffer1998} as well as in financial
time series preceding crashes \cite{riskcrash,Nasdaq,QuantFin1,QuantFin2}. In
addition, general field theoretical arguments \cite{Saleur1996b}
indicate that complex exponents are to be expected generically for
out-of-equilibrium and/or quenched disordered systems. This
together with Novikov's argument suggest to revisit
log-periodicity in turbulent signals. Demonstrating unambiguously
the presence of log-periodicity and thus of DSI in turbulent
time-series would provide an important step towards a direct
demonstration of the Richardson-Kolmogorov cascade or at least of its
hierarchical imprint.

The present paper presents the results of such an investigation on
the log-periodic oscillations of the energy dissipation rate in
3-dimensional fully developed turbulence. In section 2, we explain
where lie the difficulties of this enterprise and what are the
classical traps of standard averaging methods. We stress in
particular that standard ensemble averaging destroy the putative
log-periodic undulations in the limit of large sample sizes. To
address this problem, we explain our new ``canonical'' averaging
methodology for extracting a possible log-periodic signal in
turbulence time series and test it on synthetic examples. In
section 3, we present the result of our analysis for different
orders of the moments of the energy dissipation rate. In
particular, we discuss the sensitivity of the results with respect
to the filtering parameters of our technique. Appendix A presents
a study of the dependence of the bandwidth selectivity of the
Savitsky-Golay filter as a function of its parameters $M$ (order
of polynomial fit) and $2N_L+1$ (number of points) using the
smoothing procedure. Appendix B proposes a simple multifractal
model accounting for the observation that the same set of
log-frequencies, within numerical accuracy, are found
independently of the moment orders varying from $q=1$ to $19$.

\section{``Canonical'' averaging of undulations with noisy phases}

\subsection{Characterization of the effect of averaging on the detection
of log-periodicity}

If discrete scale invariance (DSI) exists in turbulence, this
should be reflected in the existence of a hierarchy of
characteristic scales $\ell, \ell/\gamma, \ell/\gamma^2,...$ where
$\gamma$ is a preferred scaling ratio and $\ell$ is a macroscopic scale.

Actually, this naive picture has to be revised to account for
dynamical effects. If it exists, let us imagine that the
energy cascade starts over some region at some instant from some
scale $\ell_a$ and then proceeds down the scales over a few
generations of the hierarchy.
It will probably be interrupted by various competing processes that break
down the localness of the interactions due for instance to the
influence of very
thin shear layers \cite{Kraichnan58} or slender vortex filaments
\cite{Kraichnan59,Couderfil}. Later, at another time
in some other region, another energy cascade may start
from some other scale $\ell_b$ and then proceeds downwards over a
different number of generations of the hierarchy. If this picture has
some element of truth, there is an immediate consequence with respect to the
detectability of log-periodicity associated with these transient cascades:
\begin{enumerate}
\item such log-periodic undulations will be transient in a long time series;
\item different burst of log-periodic oscillations will have different phases.
\end{enumerate}
Indeed, consider a pure log-periodic signal \be S(r) = \cos (2 \pi
f \ln r -\psi_a) \label{coslog} \ee of a structure function or
moment as a function of scale $r$, where \be f=1/\ln \gamma
\label{deflogfreq} \ee is the log-frequency. The phase $\psi_a$
can be redefined as $\psi_a = 2 \pi f \ln \ell_a$ such that the
undulation becomes $\cos (2 \pi f \ln (r/\ell_a))$. The values
$r_0=\ell_a, r_1=\ell_a/\gamma, r_2=\ell_a/\gamma^2,...$ of the
scale $r$ corresponding to the maxima of the cosine function
provide the discrete scales of the underlying cascade, which are
all proportional to the ``root'' length scale $\ell_a$. If a
second cascade is triggered from a slightly different scale
$\ell_b$, the resulting oscillation $\cos (2 \pi f \ln r -\psi_b)$
with the same log-frequency will be out-of-phase compared to the
first one with a phase shift $\psi_a - \psi_b = 2 \pi f \ln
(\ell_a/\ell_b)$. This scenario suggests that, while the physics
of the cascade may be universal with an universal scaling factor
$\gamma$, the phases are sensitively dependent on the specific
scale from which a given transient cascade nucleates. As the
cascade may be dynamically triggered from a variety of scales
$\ell_a$ from the integral scale and deep in the inertial range,
the corresponding phases are expected to be non-universal.
Averaging a signal over several such transient cascades will thus
wash out the information by ``destructive interferences'' of the
log-periodic oscillations, giving them the appearance of noise.

A similar situation has already been documented in other systems.
Numerical simulations
on Laplacian growth models \cite{Sornette1996a,Huang1997,Johansen1998}
and renormalization group calculations
\cite{Saleur1996b} have taught us that the presence of noise or disorder
modifies the phase in the log-periodic oscillations in a sample
specific way leading to a ``destructive interference'' upon
averaging. As sample-to-sample log-periodic undulations are
destroyed by averaging over a long record, this is related to the
self-averaging property \cite{Binder1986,Pazmandi1997,Wiseman1998b}.

\subsection{Existing methods for detecting log-periodicity \label{listmethods}}

Up to now, four different methods have attempted to improve on
standard averaging techniques which, in our context,
may ``throw away the baby with the bath''.
\begin{enumerate}

\item A first method consists in fitting data with a power law decorated
by the log-periodic corrections as done to improve the
determination of the time of occurrence of rupture
\cite{Anifrani1995,critrup}, of earthquakes
\cite{Sornette1995,Saleur1996a,Johansen1996,Johansen2000a} and of
financial crashes \cite{sorjohbouch,riskcrash,Nasdaq,QuantFin2}.
This parametric approach has the drawback of becoming unstable
when too much noise is present.

\item The second idea is to avoid performing any average altogether
and, studying the
main log-frequency of each realization,
to construct their distribution over the whole set of realizations. This was
used in an early attempt to demonstrate the presence of DSI structure
in growing diffusion-limited-aggregation clusters \cite{Johansen1996}.

\item The third non-parametric approach uses the fact that a
periodogram obtained
with the Lomb method \cite{Press1996} (see also below for more details)
gives the power spectrum as a function of log-frequency
independently of the phase of the undulations. As a consequence, if there
is a genuine peak, even with a lot of noise,
performing a Lomb periodogram for each sample and then averaging the Lomb
periodograms will tend to make the genuine peak stand out of the noise
if there are sufficiently many independent samples. This is the method
used for the analysis of 2-D freely decaying turbulence \cite{Johansen2000b}
already mentioned.

\item Ref.~\cite{Johansen1998} have adapted to the problem of log-periodic
oscillation
the novel so-called ``canonical'' averaging scheme proposed in
\cite{Pazmandi1997}
based on the determination of a realization-dependent effective
critical point obtained from, {\it e.g.}, a maximum susceptibility criterion.
Conceptually, the method re-sets the nucleus scale $\ell_a$ of the cascade in
different realizations to approximately the same value and then only afterwards
perform the averaging. The non-parametric method has been tested successfully
on diffusion limited aggregation clusters and on a model of
rupture \cite{Johansen1998}. For the problem of
diffusion-limited aggregation, ``canonical'' averaging amounts to
average at fixed numbers of particles per cluster. In the rupture problem,
a maximum rate of released energy is used to determine
the position of the realization-specific critical control parameter
which is then used
to perform an average over many different realizations
at fixed reduced distance from the critical rupture.
Ref.~\cite{Sornette1998b} suggested to adapt this idea to turbulence by
dividing a long turbulence
time series into intervals of length equal to a turn-over time
and rephasing the structure functions estimated over each such time
interval by identifying a scale
at which the dissipation rate is locally the largest.

\end{enumerate}

\subsection{Our new improved ``canonical'' averaging procedure
\label{ourprocedure}}

Here, we present a novel extension of the fourth method that turns
out to perform extremely well. We now describe and illustrate it
on simple synthetic data before turning to the full-fledge
analysis of a long turbulent time series. Since our goal is to detect
log-periodicity
in the structure function for which the natural variable is the logarithm
of the scale, we can rephrase the problem into one of detecting a
periodic undulation in a noisy time series.

Consider for the sake of presentation, a signal similar to
(\ref{coslog}). Posing $t \equiv \ln r$, we have the following
``time'' series \be y(t) = \cos (2 \pi f t + \psi(t))~,
\label{coslogtt} \ee where the phase $\psi(t)$ is now time varying
to account for the variation of the scale of the nucleus of the
cascade described above. Here, we do not show tests with additive
noise as this standard situation was studied in
\cite{Press1996,Huang2000b} and is relatively easily addressed
with the Lomb periodogram spectral analysis (see also below). We
thus focus our attention on the most difficult situation of random
phases.

To be specific,
let us assume that $\psi(t)$ undergoes a random walk
\be
\psi(t)= \psi(t-dt) + 2 \pi \sqrt{f}~ A ~{\rm ran}(t)~,
\label{psifjnal}
\ee
where ${\rm ran}(t)$ is a Gaussian random number of zero mean and of
variance $dt$.
With this parameterization (\ref{psifjnal}), over $N$ periods $1/f$
of the deterministic
component, the phase wanders by an amplitude of the order of $2 \pi A
\sqrt{N}$ which
is rapidly larger than $2 \pi$ if $A$ is comparable to $1$ as we discuss here.
This random phase modulation can also be seen as a multiplicative noise
acting on the signal $(\cos (2 \pi f t), \sin (2 \pi f t))$
in the complex plane since $\cos (2 \pi f t + \psi(t))=
\cos \psi ~\cos (2 \pi f t)  - \sin \psi ~\sin (2 \pi f t)$.
In the absence of the deterministic signal $2 \pi f t $ in the cosine in
expression (\ref{coslogtt}), the random process $y(t)$ has a
correlation function
which can be evaluated exactly: $\langle \cos (\psi(t+\tau)) \cos
(\psi(t)) \rangle
-\langle \cos (\psi(t+\tau)) \rangle \langle \cos (\psi(t)) \rangle =
e^{-2\pi^2 A^2 \tau}$ with a characteristic correlation time $1/2\pi^2 A^2$.

\begin{figure}
\begin{center}
\epsfig{file=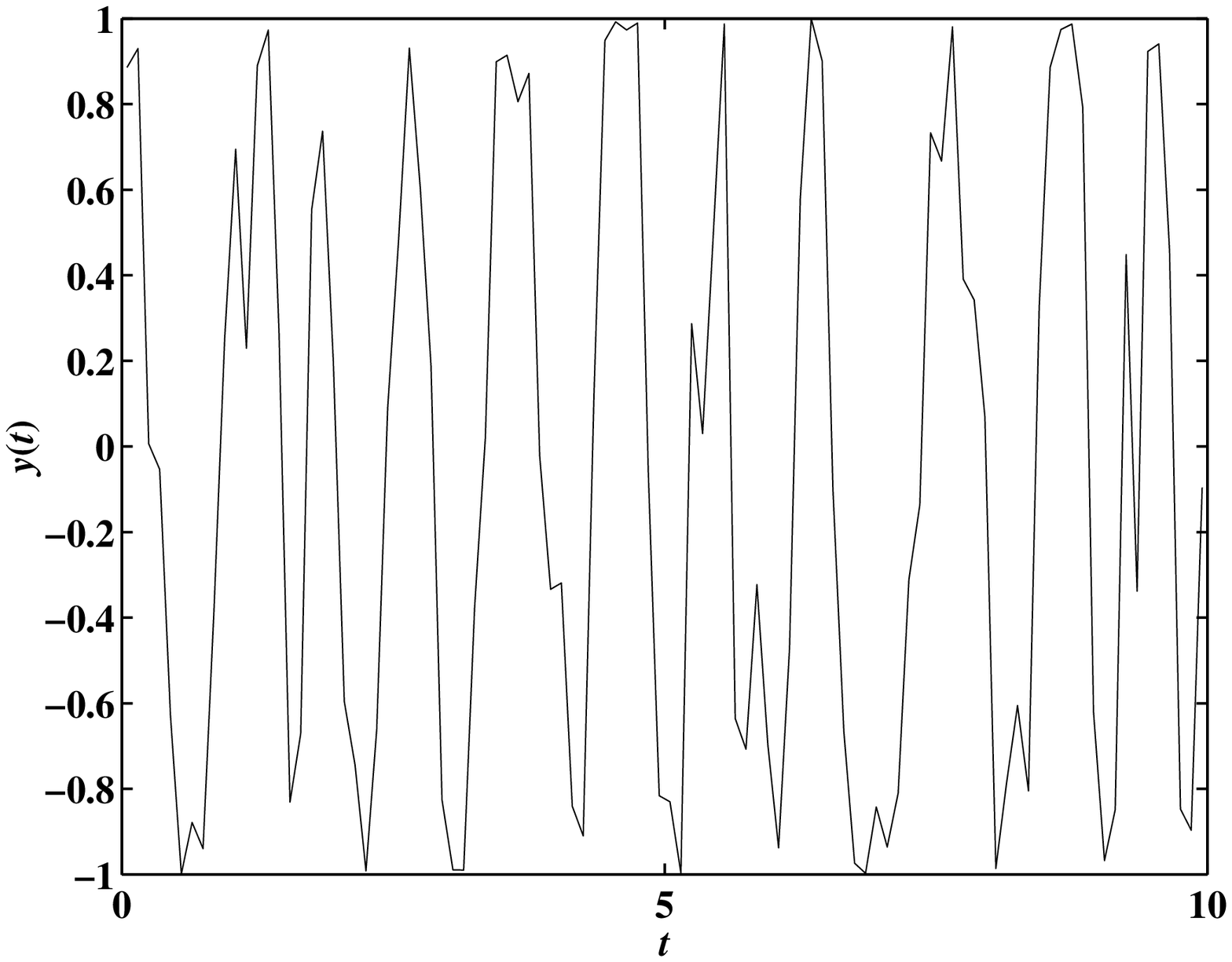,width=12cm, height=8cm}
\end{center}
\caption{One realization with $100$ points of the time series
(\ref{coslogtt},\ref{psifjnal}) with $f=1, A=0.3$ generated
numerically with $dt=0.1$ in the time interval $[0, 10]$.
}
\label{figsigneta1}
\end{figure}

\begin{figure}
\begin{center}
\epsfig{file=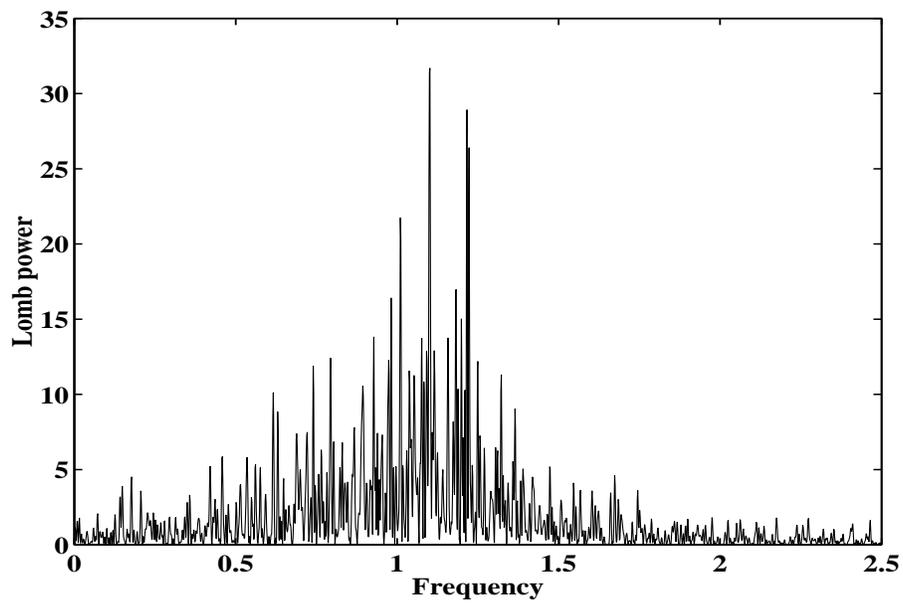,width=12cm, height=8cm}
\end{center}
\caption{Standard spectral analysis (Fourier spectrum) of a long time series
(\ref{coslogtt},\ref{psifjnal}) with $f=1, A=0.3$ generated
numerically with $dt=0.1$ in the time interval  $[0,200]$
with $2000$ points.
}
\label{signeta1stanfourier}
\end{figure}

Figure \ref{figsigneta1} shows one realization with $100$ points of
the time series
(\ref{coslogtt},\ref{psifjnal}) with $f=1, A=0.3$ generated
numerically with $dt=0.1$ in the time interval $[0, 10]$. This corresponds
to a correlation time $1/2\pi^2 A^2 = 0.44$.
This value $A=0.3$ thus corresponds to a very strong phase disorder, so strong
that the time series appears very random.
Figure \ref{signeta1stanfourier}
shows a standard spectral analysis (Fourier spectrum) of a long time series
with $2000$ points in the time interval $[0,200]$. It is clear that the
random walk of the phase $\psi(t)$ destroys completely the
information of the existence
of an underlying deterministic component at the frequency $f=1$: the
figure shows
a completely noisy spectrum.

\begin{figure}
\begin{center}
\epsfig{file=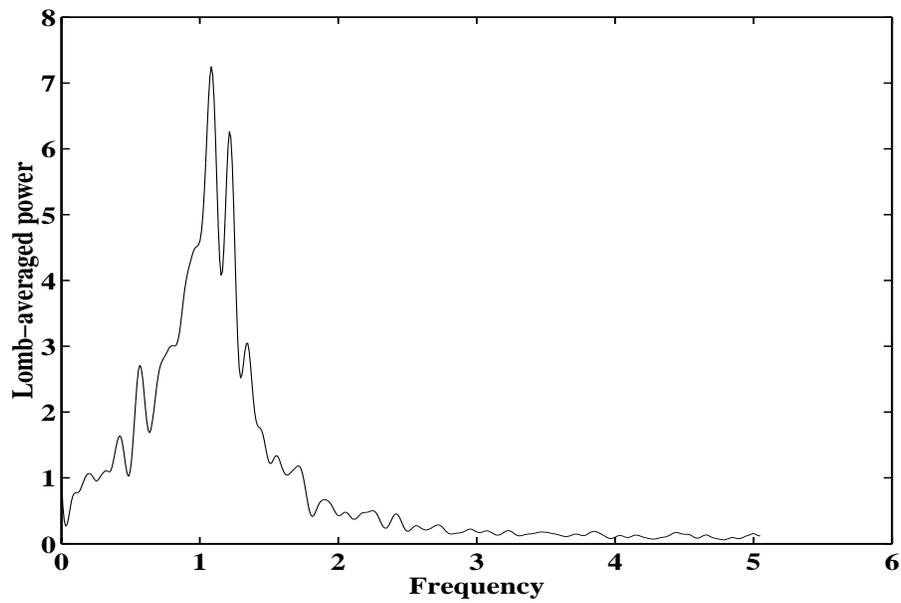,width=12cm, height=8cm}
\end{center}
\caption{Result of the third method listed in section
\ref{listmethods} applied to the time series of $2000$ points
generating to obtain the spectrum shown in Fig.
\ref{signeta1stanfourier}. We divide the long time series with
$2000$ points in $20$ contiguous time series of $100$ points each
that are similar to that represented in Fig. \ref{figsigneta1}. We
perform on each of these $20$ time series a Lomb periodogram
analysis and then average these $20$ Lomb periodograms. }
\label{signeta1Lombav}
\end{figure}

Figure \ref{signeta1Lombav} shows the result of the third method
listed in section \ref{listmethods}: we take the long time series
with $2000$ points in the time interval $[0,200]$ and divide it in
$20$ contiguous time series of $100$ points each that are similar
to that represented in Fig. \ref{figsigneta1}. We perform on each
of these $20$ time series a Lomb periodogram analysis and then
average these $20$ Lomb periodograms. The Lomb periodogram
analysis performs local least-square fits of the data by sinusoids
centered on each data point of the time series \cite{Scarg1982}.
The Lomb periodogram usually out-performs significantly other
spectral methods for unevenly sampled and relatively short signals
\cite{Scarg1982,Horne1986,Press1996,Zhou2001b}. Figure
\ref{signeta1Lombav} provides already a significant improvement
over the standard Fourier analysis of Fig.
\ref{signeta1stanfourier}, as a quite broad but rather clear peak
centered approximately on $f=1$ can be distinguished. However, its
statistical significance is not convincing since the hypothesis
that it could result just from random occurrences can not be
rejected by standard statistical tests \cite{Press1996,Zhou2001b}.

\begin{figure}
\begin{center}
\epsfig{file=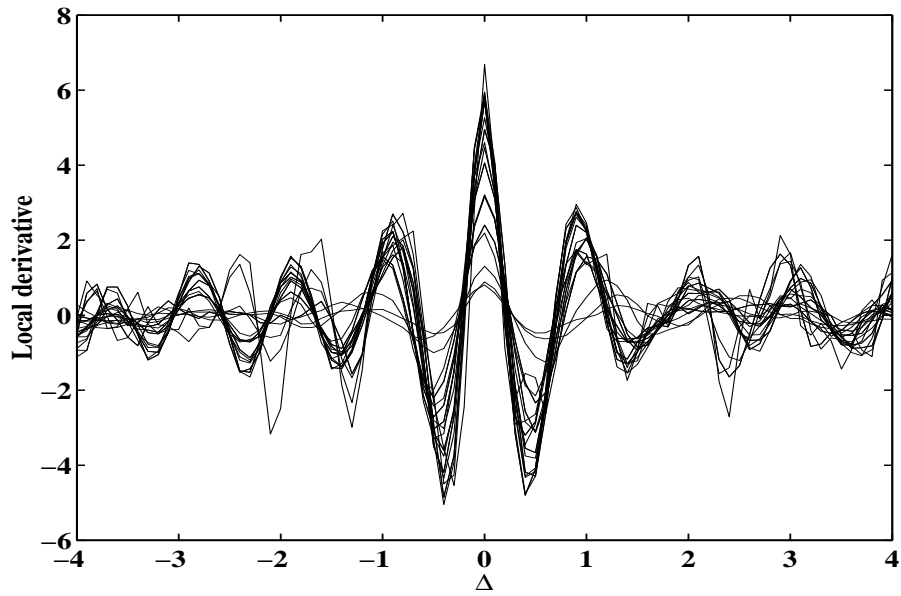,width=12cm, height=8cm}
\end{center}
\caption{For a given $N_L$ and $M$, we perform the average of the $20$
time series derivatives described in the text. Varying $N_L$ between
$5$ and $10$
and $M$ between $4$ and $7$, both by unit increments, provides
a total of $24$ averaged time series derivatives that are
shown here.
}
\label{signeta1rephas}
\end{figure}

\begin{figure}
\begin{center}
\epsfig{file=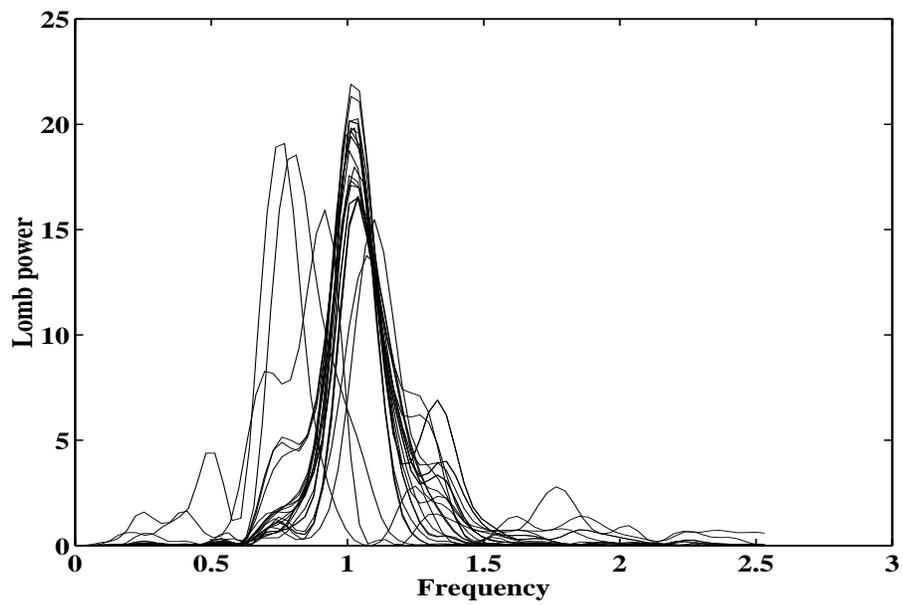,width=12cm, height=8cm}
\end{center}
\caption{For each of the $24$ averaged time series derivatives
represented in Fig. \ref{signeta1rephas}, we show the
corresponding Lomb periodogram spectrum. }
\label{signeta1lombavto}
\end{figure}

\begin{figure}
\begin{center}
\epsfig{file=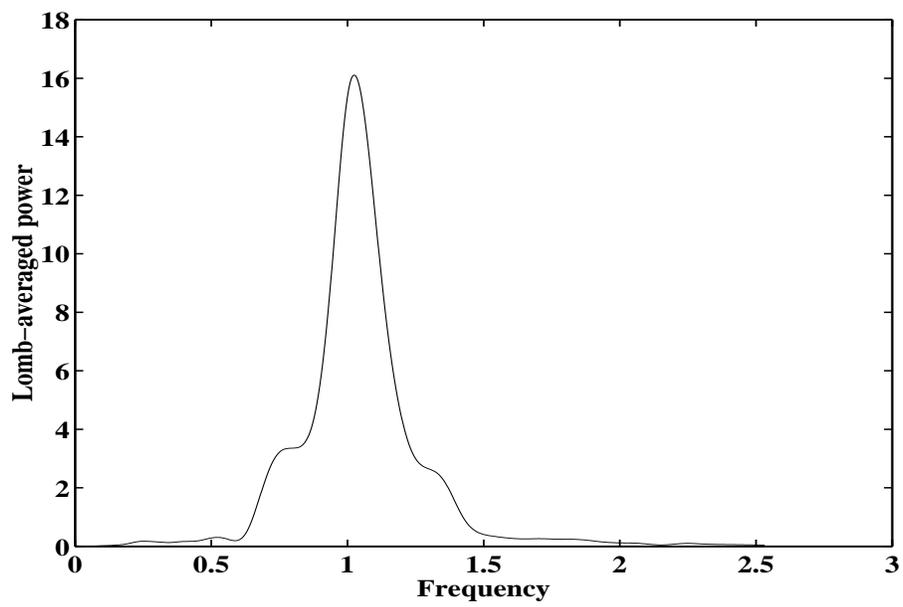,width=12cm, height=8cm}
\end{center}
\caption{Average over the $24$ Lomb spectra shown in Fig.
\ref{signeta1lombavto}. } \label{signeta1rephasave}
\end{figure}

Our new method consists in the following steps.
\begin{enumerate}
\item Rather than analyzing the long time series
of $2000$ points, we divide it again in $20$
contiguous time series of $100$ points each: $[1, 100]; [101, 200]; ...
[1+n\times 100, (n+1)\times 100];...; [1901, 2000]$.
These two numbers $20$ and $100$ are
rather arbitrary and can be modified by and large. We need however
a sufficient number of times series to average over as discussed
below and a sufficient number of points in each time series to
carry at least a few oscillations.

\item On each sub-time series of $100$ points, we apply
the Savitsky-Golay filter \cite{Press1996} over a running window
of size $N_L+N_R+1$ (with $N_L$ points to the left and $N_R$
points to the right of a running point) by fitting a polynomial of
order $M$. We take $N_L=N_R$ (symmetric filter) and vary $N_L$
between $5$ and $10$ and $M$ between $4$ and $7$, both by unit
increments. This provides a total of $24$ filtered versions of
each sub-time series of $100$ points. The Savitsky-Golay filter
has the advantage of providing smoothing without loss of
resolution and of avoiding the distortion of the moments of the
signal of all orders up to the order of the polynomial. We vary
$N_L$ and $M$ over rather broad intervals to check for the
sensitivity of this filtering process and to avoid the occurrence
of spurious frequencies that may be created by the filter (see
Appendix A).

\item For each of the $24$ filtered versions of each sub-time series
of $100$ points,
we use its local polynomial representation to calculate analytically its
local derivative by simply differentiation the polynomial associated with
each point along the time series.
The step constructs an observable which is sensitive to the phase.

\item For each time series of the derivative, we identify the time
$t_{\rm max}$
of the local maximum closest to its middle point. This is analogous to
identifying a local maximum of the control parameter of the ``canonical''
averaging method described above \cite{Pazmandi1997,Johansen1998}.

\item We then translate the origin
of time of each derivative time series to this time $t_{\rm max}$.
This is the ``re-phasing'' step.

\item For a given $N_L$ and $M$, we perform the average of the $20$
time series derivatives. Varying $N_L$ between $5$ and $10$ and
$M$ between $4$ and $7$, both by unit increments, provides a total
of $24$ averaged time series derivatives that are shown in Fig.
\ref{signeta1rephas}. This is the ``canonical'' averaging step.
The figure is not unlike the local correlation structure $g(r)$
obtained in the probing of the local order of liquids around a
representative molecule. In particular, the oscillations decay in
amplitude as a function of the distance to the central peak. Note
that our choice of rephasing the local maximum closest to its
middle point (and not other point) of each time series provides
the best choice in order to observe a maximum number of
oscillations.

\item For each of the $24$ averaged time series derivatives, we
perform a Lomb periodogram analysis in order to extract the
most probable frequency of each signal. Figure \ref{signeta1lombavto}
shows the corresponding $24$ Lomb spectra. A significant fraction of
the $24$ Lomb spectra
exhibit a very strong and statistically significant thin peak at the correct
frequency $f=1 \pm 0.05$. Other spurious peaks are also present:
they result from the interplay between Savitsky-Golay smoothing procedure and
the phase noise.

\item To confirm that the information on the existence of the deterministic
signal at the frequency $f=1$ is present in each of the $24$ Lomb
spectra, Fig. \ref{signeta1rephasave} shows the average over these
$24$ Lomb spectra in Fig. \ref{signeta1lombavto}. A very clear
peak at $f=1.03 \pm 0.1$ is obtained with an amplitude twice as
big as the background.

\end{enumerate}

We have presented here our new method on an extreme case of phase
disorder. For smaller disorder, for instance already with $A=0.1$, the signal
obtained with our method is overwhelming while the direct Fourier
spectral analysis or even the third method of direct Lomb averaging
provide still
unconclusive proof of the existence of a deterministic frequency.

We now present the application of this technique for the
detection of log-periodicity in the moments of the energy
dissipation rate.

\section{Detection of log-periodicity in the moments of the energy
dissipation rate}

\subsection{Standard preliminary tests on the experimental data
\label{sec:expt}}

\begin{figure}
\begin{center}
\epsfig{file=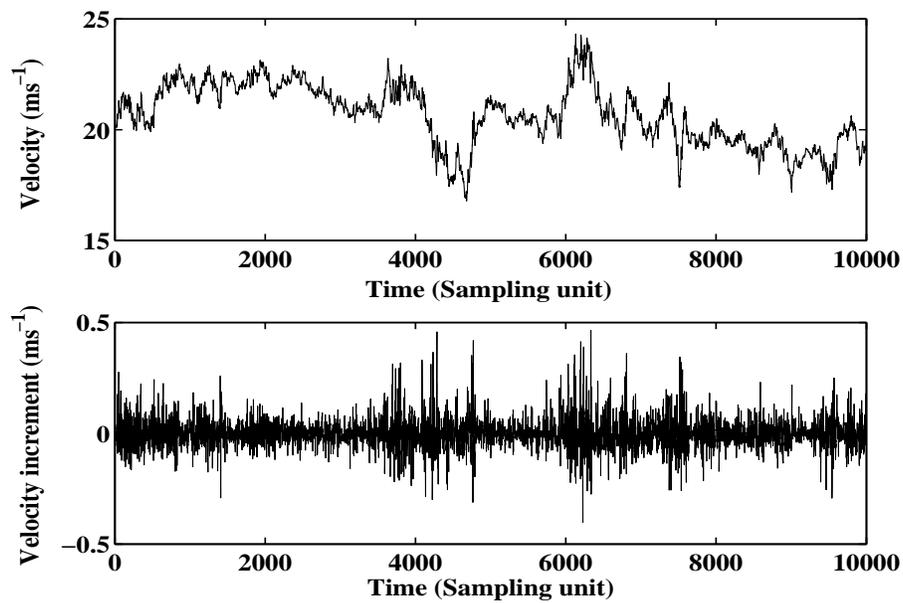,width=12cm, height=8cm}
\end{center}
\caption{(a) A typical streamwise velocity time series recorded by hot-wire
anemometry in the S1 ONERA wind tunnel. (b) Corresponding time series
of the velocity
increments.}
\label{Fig:Velocity}
\end{figure}

Very good quality high-Reynolds turbulence data have been collected
at the S1 ONERA wind tunnel by the Grenoble group from LEGI
\cite{Anselmet1984}.
We use the longitudinal velocity data obtained from this group to check
for the possible presence of log-periodic undulation in moments
of the energy dissipation rate. Figure \ref{Fig:Velocity} shows a
typical sample of the time series of the
streamwise component of the flow velocity, denoted as $v$ hereafter
in this work.

The mean velocity of the flow is approximately $\langle{v}\rangle =
20 \mathtt{m s^{-1}}$ (compressive effects are thus negligible).
The root-mean-square velocity fluctuations is
$v_{\mathtt{rms}} = 1.7 \mathtt{m s^{-1}}$, leading to a turbulence
intensity equal to $I = {v_{\mathtt{rms}}} /
{\langle{v}\rangle} = 0.0826$. This is sufficiently small to allow for
the use of Taylor's frozen flow hypothesis. The integral scale is
approximately $4 \mathtt{m}$ but is difficult to estimate precisely
as the turbulent
flow is neither isotropic nor homogeneous at these large scales.

The Kolmogorov microscale $\eta$ is given by \cite{Meneveau1991}
$\eta = \left[\frac{\nu^2 \langle{v}\rangle^2}{15 \langle(\partial
v/\partial t)^2\rangle }\right]^{1/4} = 0.195 \mathtt{mm}$,
where $\nu = 1.5 \times 10^{-5} \mathtt{m^2 s^{-1}}$ is the
kinematic viscosity of air. $\partial v/\partial t$ is evaluated by
its discrete approximation with a time step increment
$\partial t = 3.5466 \times 10^{-5} \mathtt{s}$ corresponding to the spatial
resolution $\delta_\ell = 0.72 \mathtt{mm}$ divided
by $\langle{v}\rangle$.

The Taylor scale is given by \cite{Meneveau1991}
$\lambda =\frac{\langle{v}\rangle
v_{\mathtt{rms}}}{\langle (\partial v/\partial t)^2 \rangle^{1/2}}
=16.6 \mathtt{mm}$. The Taylor scale is thus about $85$ times the
Kolmogorov scale. The Taylor-scale
Reynolds number is $Re_\lambda = \frac{v_{\mathtt{rms}}\lambda}{\nu} = 2000$.
This number is actually not constant along the whole data set and
fluctuates by about $20\%$.

We have analyzed a total number of about ${\cal N}=1.4 \times 10^7$
data points provided in
$200$ segments of $2^{16}=65536$ data points. In order to implement
our detection
method presented in section \ref{ourprocedure}, we have
re-partitioned the total data set
of ${\cal N}$ points
into $N$ ``records'' of length $L$. In the sequel, we present results for
$N=100$ records of length
$L=2^{17} \approx 130,000$ data points ($N \times L={\cal N}$) and
for $N=20$ records of $L=5 \times 2^{17} \approx 655,000$ data points.
Having $N=100$ (or $N=20$ respectively) such
records will allow us to test for the reproductivity and stationarity
of the results.
Each of the $N$ records of length $L$ is then itself subdivided
into ``samples'' of $2^{11}=2,048$ data points.
We thus have $64$ (respectively $320$)
samples per record for $L=2^{17}$ (respectively for $L=5 \times 2^{17}$).
Taking an integral scale of
$\ell_0=4 \mathtt{m}$ with one point per $0.72 \mathtt{mm}$,
a complete turnover covers approximately $2 \ell_0/\delta_\ell \approx 10^5$
data points. Each sample thus covers a finite fraction ($\approx
1/5$) of a turn-over time.

We have checked that the turbulent velocity time series recovers
the standard scaling laws previously reported in the literature
with similar quality. In particular, we have verified the validity
of the power-law scaling $E(k) \sim k^{-\beta}$ with an exponent
$\beta$ very close to $\frac{5}{3}$ over a range more than two
decades, similar to Fig. 5.4 of \cite{Frisch1996} provided by Y.
Gagne and M. Marchand on a similar data set from the same
experimental group. Similarly, we have checked carefully the
determination of the inertial range by combining the scaling
ranges of several velocity structure functions \cite{Gagne1987}
(see also Fig. 8.6 of \cite{Frisch1996}). Conservatively, we are
led to a well-defined inertial range $60 \leq \ell/\eta \leq
2000$.

\subsection{Moments of the energy dissipation rate}

Using Taylor's hypothesis which replaces a spatial variation of the fluid
velocity by a temporal variation measured at a fixed location,
the rate of kinetic energy dissipation at position $i$ is
\begin{equation}
\epsilon_i \sim \left[ \left(v_{i+1} - v_i \right) / \delta_\ell \right] ^2~,
\label{Eq:epsilon}
\end{equation}
where $\delta_\ell$ is the resolution (translated in spatial
scale) of the measurements. The total dissipation rate $E_\ell$ in a
spatial domain $\Omega$ of size $\ell$ is
\begin{equation}
E_\ell = \delta_\ell~\sum_{\Omega_i\in\Omega} \epsilon_i ~,
\label{Eq:Eell}
\end{equation}
where $\Omega_i$ is the $i$th sub-piece of linear dimension $\delta_\ell$
in $\Omega$, such that $\Omega_i \cap \Omega_j =
\Phi$ for $i \neq j$ and $\bigcup_i \Omega_i =
\Omega$. We use normalized energy dissipation
$E_\ell/E_L$, where $L$ is the size of the system, and normalized
scale $\ell/\eta$.

We study the $q$-th
moment of the normalized energy dissipation rate
\begin{equation}
M_q(\ell) = \sum \left(E_\ell/E_L\right)^q~,
\label{Eq:Mq}
\end{equation}
as a function of the scale $\ell$. The summation in (\ref{Eq:Mq})
is performed over all domains of size $\ell$ within
the system of size $L$. The signature of multifractal scaling is that
the scaling law
\begin{equation}
M_q(\ell) \sim \left(\ell/\eta\right)^{\tau(q)}, \label{Eq:tau}
\end{equation}
is expected to hold in the inertial range with moment
exponents $\tau(q)$ that are non-linear functions of $q$.
Using standard arguments
\cite{ParisiFrisch,Halsey1986,Frisch1996}, the multifractal spectrum
$f(\alpha)$
is the Legendre transform of $\tau(q)$.
The validity of these
power laws (\ref{Eq:tau}) and multifractality
have been verified with a good accuracy in experiments
\cite{Meneveau1991,Frisch1996}. We have reproduced these results reliably.
As already mentioned, one can discern sometimes some small noisy oscillations,
which are suggestive of log-periodicity.

\subsection{Canonical averaging of the moment exponents \label{secondmeth}}

In order to test for the possible existence of DSI, we apply the
methodology explained in section \ref{ourprocedure} to each of
the $N$ records of length $L=2^{17}$. We estimate the moments $M_q(\ell) $
defined by (\ref{Eq:tau}) in the range $\ell/\eta \in [216,1840]$
which is well within the inertial range for each sample. The lower
boundary $216$ is $3.6$ times larger than the lower bound $60$ of the
inertial range previously discussed, in order to avoid
the effect of very small scales that make the signal more noisy.

\begin{enumerate}
\item Given a record of length $L=2^{17}$ (resp. $L=5 \times 2^{17}$),
we cut it into $64$
(resp. $320$) samples of size $2^{11}$ points.

\item We then calculate the local moment exponent
$\tau \left(q, \ln(\ell/\eta) \right)$ as the logarithmic
derivative of $M_q(\ell)$ with respect to $\ell/\eta$:
\begin{equation}
\tau \left(q, \ln(\ell/\eta) \right) =
\frac{d\ln(M_q(\ell))}{d\ln(\ell/\eta)}~.
\label{Eq:tauqell}
\end{equation}

\item For each sample, we then identify the sample-dependent reduced
scale $\ell_c/\eta$
closest to the central point of the scale-interval of the analysis
at which $\tau \left(q, \ln(\ell/\eta) \right)$ is maximum.

\item We rephase all functions $\tau \left(q, \ln(\ell/\eta) \right)$
of the variable $\ln(\ell/\eta)$ of each sample so that their
$\ell_c/\eta$ coincide. This allows us to define the reduced
variable $\Delta \equiv \ln(\ell/\eta)- \ln(\ell_c/\eta)$ such
that all functions $\tau \left(q, \Delta \right)$ are rephased at
$\Delta =0$.

\item We then average these rephased functions $\tau \left(q, \Delta \right)$.
This defines the function (local average exponent) \be D(\Delta) =
\langle \tau \left(q, \Delta \right) \rangle~,
\label{localderjmsl} \ee where the average is performed over the
different samples in one record at fixed $\Delta = \ln(\ell/\eta)-
\ln(\ell_c/\eta)$ (where $\ln(\ell_c/\eta)$ vary from sample to
sample and $\ln(\ell/\eta)$ is adjusted accordingly).

\item We then perform several comparative analysis to check the
quality of the results
discussed in the following.
\end{enumerate}

\subsection{Results \label{subsec:Expt}}

First, let us stress that our results presented below have been found
to yield the same set of log-frequencies, within
numerical accuracy, independently of the moment orders varying from
$q=1$ to $19$.
Appendix B proposes a simple multifractal model explaining this observation.
The standard deviation of the log-frequencies discussed below determined
from the $N=100$ records of length $L=2^{17}$ shows that it is larger for $q=1$
and reaches a plateau at $q=3$. We have thus chosen to present the following
results for $q=3$.

\begin{figure}
\begin{center}
\epsfig{file=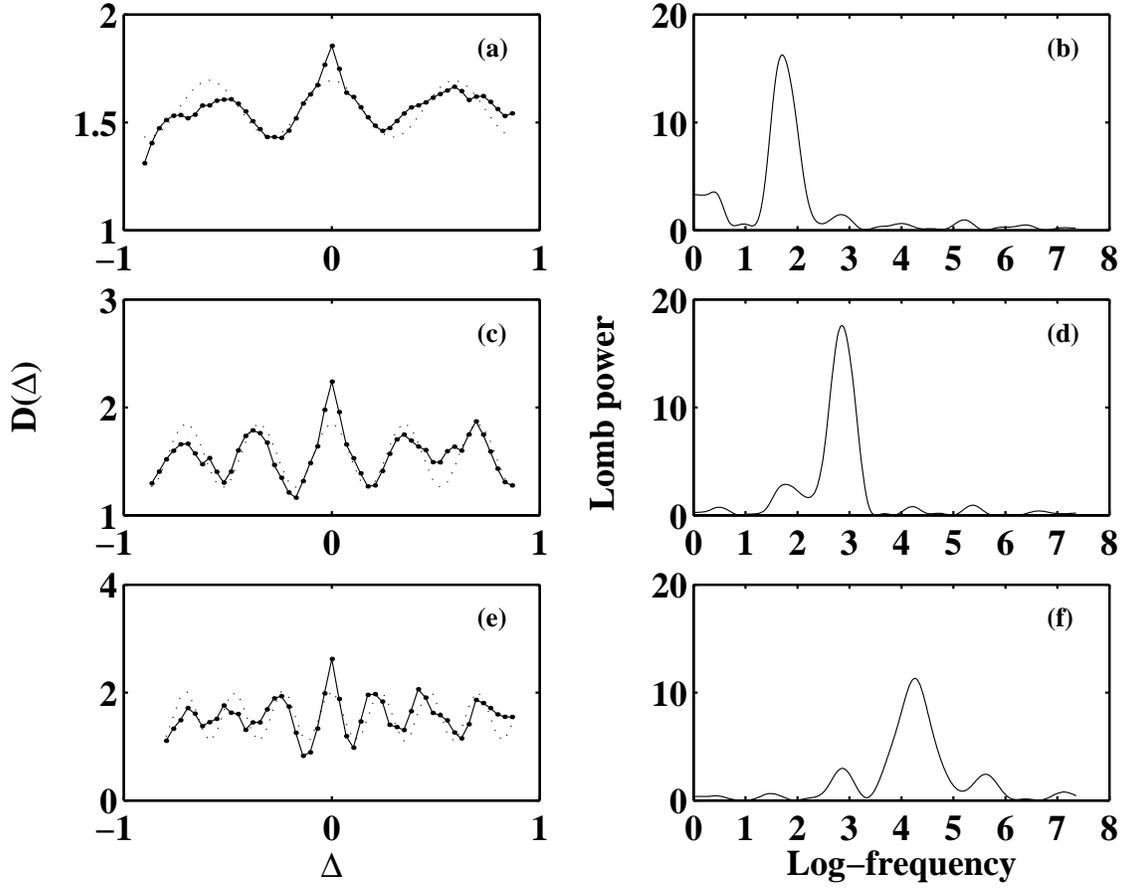,width=15cm, height=12cm}
\end{center}
\caption{Local
log-derivative $D(\Delta)$ defined in (\ref{localderjmsl}),
canonically averaged over the $64$ samples of a single record of length
$L=2^{17}$ points,
as a function of $\Delta$ (left panels) and their corresponding
Lomb periodograms (right panels) for three choices of the
Savitsky-Golay filter parameters $N_L$ and $M$.
(a-b) $M = 4$ and $N_L = 10$, (c-d) $M = 5$ and
$N_L = 9$, (e-f) $M = 7$ and $N_L = 8$.
The dotted line in the left panels are pure cosine functions
$D(\Delta) = \langle D \rangle + \sqrt{2}\sigma_D \cos(2\pi f \Delta)$.
}
\label{Fig:3Types}
\end{figure}

Figure~\ref{Fig:3Types} shows the canonically averaged local
log-derivative $D(\Delta)$ defined in (\ref{localderjmsl}) as a
function of $\Delta$ (left panels) and their corresponding Lomb
periodograms (right panels) for three choices of the
Savitsky-Golay filter parameters $N_L$ and $M$. This analysis is
performed over the $64$ samples of a single record of length
$L=2^{17}$ points.

Rather strong periodic oscillations can be observed on the local
average exponent
$D(\Delta)$, which results in the high Lomb peaks in the right
panels which have good statistical significance \cite{Press1996,Zhou2001b}.
As expected, the central oscillation at $\Delta=0$ is the largest and
the undulations
on both side decay as a result of the progressively increasing destructive
interference due to noise.
One can also note a slight left-right asymmetry that we attribute to the
stronger noise affecting the small scales
($\Delta<0$) compared to the large scales ($\Delta>0$).

\begin{figure}
\begin{center}
\epsfig{file=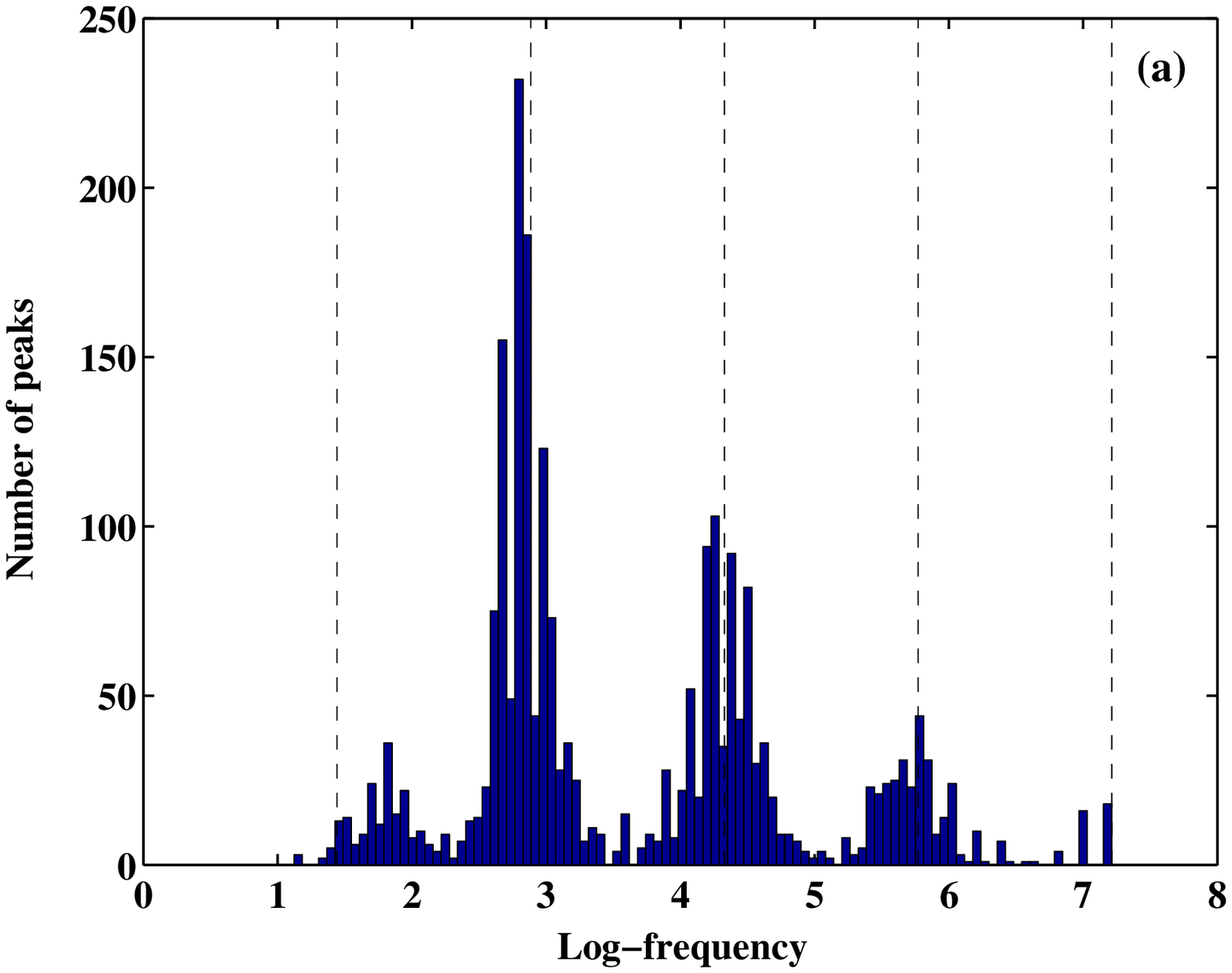,width=12cm, height=9cm}
\epsfig{file=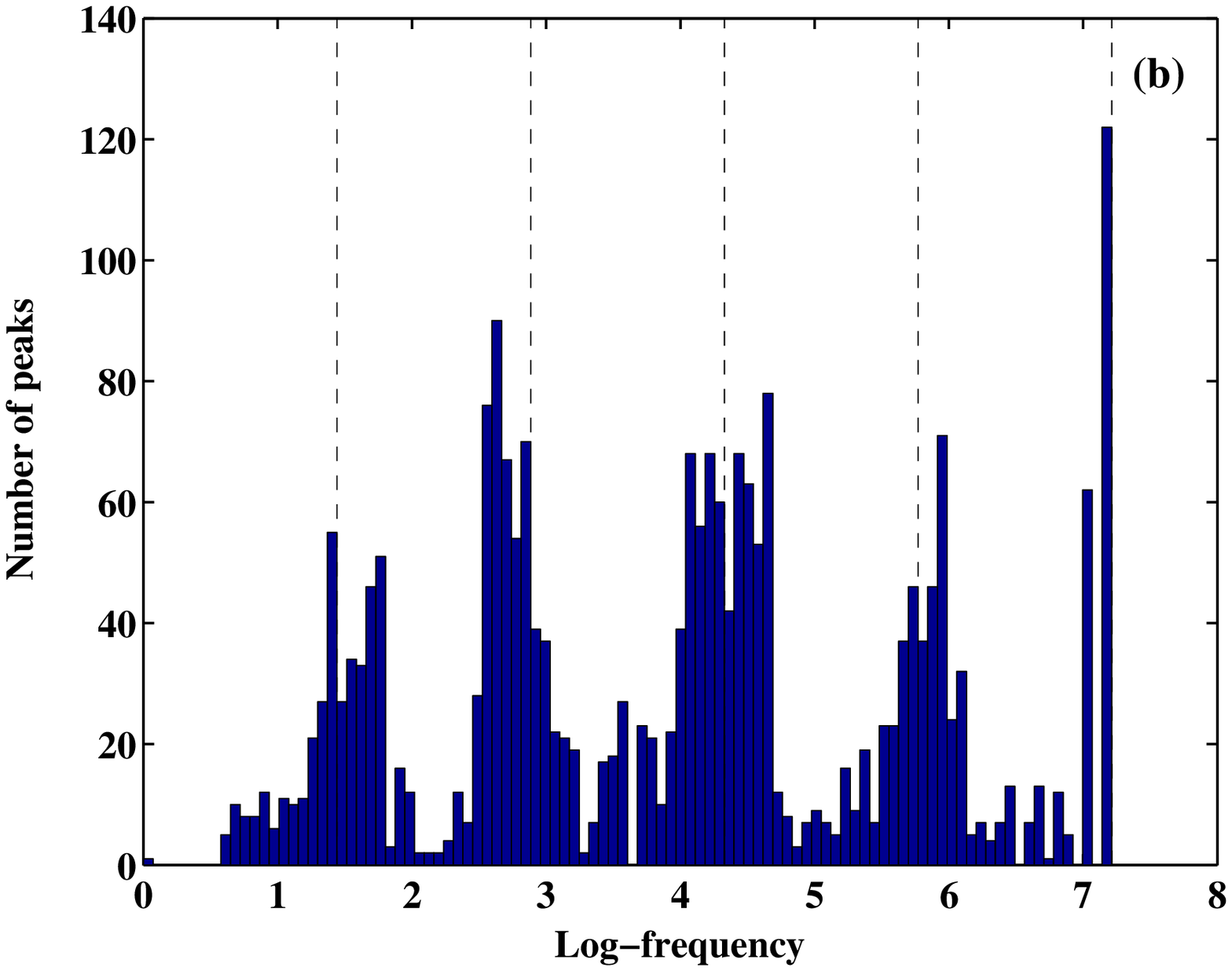,width=12cm, height=9cm}
\end{center}
\caption{Histograms of all log-frequencies corresponding to (a)
the most significant and (b) the second most significant Lomb peaks
of all 2400 Lomb periodograms available from our
analysis of $100$ independent records by the $24$ possible
filters (spanning $N_L=5-10$ and $M=4-7$) used for each record.
The vertical dashed lines correspond to log-frequencies equal respectively to
$f_1=1.44, f_2= 2 f_1 =2.89, f_3=3 f_1=4.33, f_4=4 f_1 = 5.77$ and
$f_5 = 5 f_1 = 7.21$, which correspond to
increasing integer powers of a fundamental scale ratio exactly equal to
$\gamma = 2$, through the relationship $f_1 = 1/\ln \gamma$.}
\label{Fig:Hist}
\end{figure}

\begin{figure}
\begin{center}
\epsfig{file=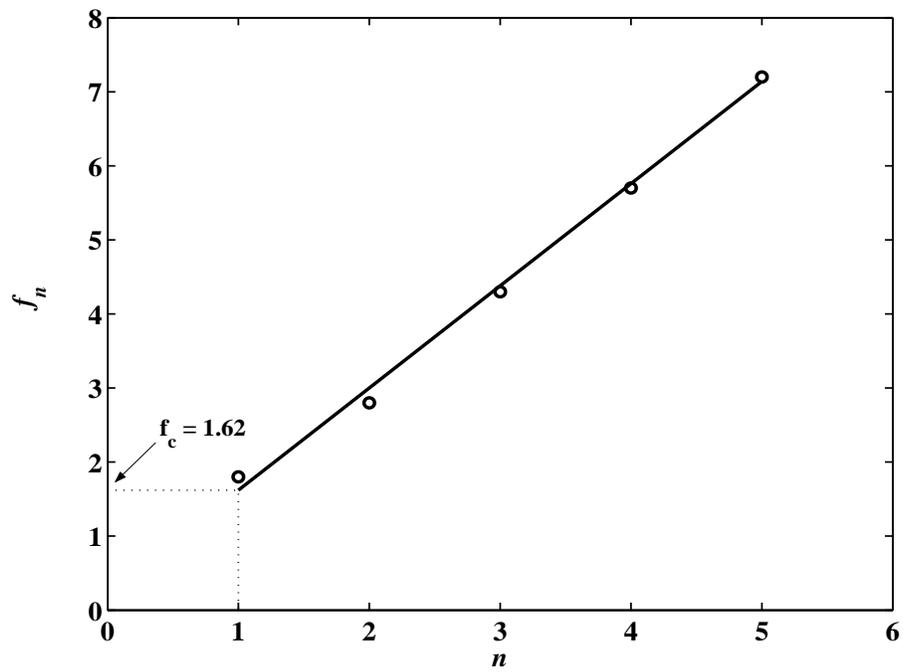,width=12cm, height=9cm}
\end{center}
\caption{Value of the frequency measured at the $n$th maximum of
the histogram shown in Fig. \ref{Fig:Hist} as a function of the
order $n$ of the maximum. The straight line is a linear fit with
the equation $f_n = n f_1$, where the adjustable parameter $f_1$
is found equal to $f_1=1.62 \pm 0.1$.} \label{Fig:HistFit}
\end{figure}

While the undulations and their Lomb spectra seem convincing,
there is a problem that seems at first sight to destroy the
evidence of a genuine log-periodicity: the periods of the
oscillations are strongly dependent upon the parameters $N_L$ and
$M$ of the Savitsky-Golay filter (see Appendix A): it would thus
seem that we are quantifying spurious oscillations created by the
massaging of the data. A similar effect has been found earlier
associated with the construction of cumulative quantities
\cite{Huang2000b}. In order to understand the possible origin of
the strong variations in significant log-frequencies as a function
of $N_L$ and $M$ and following Ref.~\cite{Sornette1996a}, we have
collected the two most significant log-frequencies (two highest
Lomb peaks) of all the $2400$ Lomb periodograms available from our
analysis of $100$ independent records by the $24$ possible filters
(spanning $N_L=5-10$ and $M=4-7$) used for each record.
Fig.~\ref{Fig:Hist}(a) (respectively (b)) shows the histogram of
the log-frequencies associated with the largest (respectively
second largest) Lomb peaks. The vertical dashed lines correspond
to log-frequencies equal respectively to $f_1=1.44, f_2= 2 f_1
=2.89, f_3=3 f_1=4.33, f_4=4 f_1 = 5.77$ and $f_5 = 5 f_1 = 7.21$.
These values correspond respectively to increasing harmonics of a
fundamental frequency $f_1 = 1/\ln \gamma$ associated with the
scale ratio $\gamma = 2$. Remarkably, similar histograms (but with
smaller amplitudes for the peaks at $f_2$ and $f_3$) are also
obtained by collecting the frequencies corresponding to the third
and fourth largest Lomb peaks among the $2400$ available Lomb
periodograms. The existence of these $5$ harmonic frequencies is
thus a very robust feature. Another evidence of the quality of the
description in terms of $5$ harmonics of a fundamental frequency,
we represent in Fig. \ref{Fig:HistFit} the value of the frequency
measured at the $n$th maximum of the histogram shown in Fig.
\ref{Fig:Hist}(a) as a function of the order $n$ of the maximum.
The straight line shows the excellent linear fit to the equation
$f_n = n f_1$, where the sole adjustable parameter $f_1$ is found
equal to $f=1.62 \pm 0.1$. This value corresponds to a prefered
scaling ratio of the cascade equal to $\gamma = 1.85 \pm 0.1$.

We are now in position for explaining the strong variations in the
position of the Lomb peaks observed in Fig. \ref{Fig:3Types}.
Since each filter has a different passing-band (see Appendix A),
for a periodic signal which has harmonics of a fundamental
frequency with strong amplitudes, these different harmonics will
be differently expressed by the different filters. This is what we
observe in Fig. \ref{Fig:3Types}. The two top panels correspond to
the filter with $(N_L=10, M=4)$ which selects preferentially the
fundamental frequency $f_1$. The two middle panels correspond to
the filter with $(N_L=9, M=5)$ which selects preferentially the
second harmonic $f_2=2 f_1$. Finally, the two bottom panels
correspond to the filter with $(N_L=8, M=7)$ which selects
preferentially the third harmonics $f_3=3 f_1$. Furthermore, the
fact that the second or third harmonics have a larger amplitude
than the fundamental frequency is not a surprise and is found for
instance in multifractal measures constructed on DSI fractal
geometries \cite{Zhou2001c}.

The limited bandwidth of each filter and their selection of
different harmonics is further quantified by Table \ref{Tb:fMNL}
which explores the range of values $N_L$ from $5$ to $10$ and $M$
from $4$ to $7$, corresponding to a total of $24$ filters. Each
filter is applied to each of the $N=100$ records of length
$L=2^{17}$ to obtain the canonically averaged $D(\Delta)$ from
which we obtain the corresponding $100$ log-frequencies as
explained in section \ref{secondmeth}. The average $\langle f
\rangle$ of these $100$ log-frequencies and their standard
deviation $\sigma_f$ are given as a function of $M$ and $N_L$ in
Table \ref{Tb:fMNL}. $\langle f \rangle$ increases with increasing
$M$ and/or decreasing $N_L$, an expected property of the
Savitsky-Golay filter (see Appendix A). This is not however where
lies the potentially useful information. To retrieve it, we
identify ten pairs $(M, N_L)$, specifically $(4,6)$, $(4,7)$,
$(4,10)$, $(5,8)$, $(5,9)$, $(5,10)$, $(6,8)$, $(6,9)$, $(6,10)$,
and $(7,8)$, whose $\sigma_f \leq 0.3$, i.e., such that there is
good confidence in the determination of a log-frequency.
Interestingly, these ten pairs give three groups of
log-frequencies: (1) eight pairs marked by astroids ($*$) have an
averaged log-frequencies in the range $2.71-2.99$; (2) one pair
marked by the star ($\star$) has $\langle f \rangle =4.37$; (3)
one pair marked by the circle ($\circ$) has $\langle f \rangle =
1.79$. These three frequency bands are in good agreement with the
values for the harmonics $f_1=1.44$, $f_2= 2 f_1 =2.89$ and $f_3=3
f_1=4.33$ already discussed.

\begin{figure}
\begin{center}
\epsfig{file=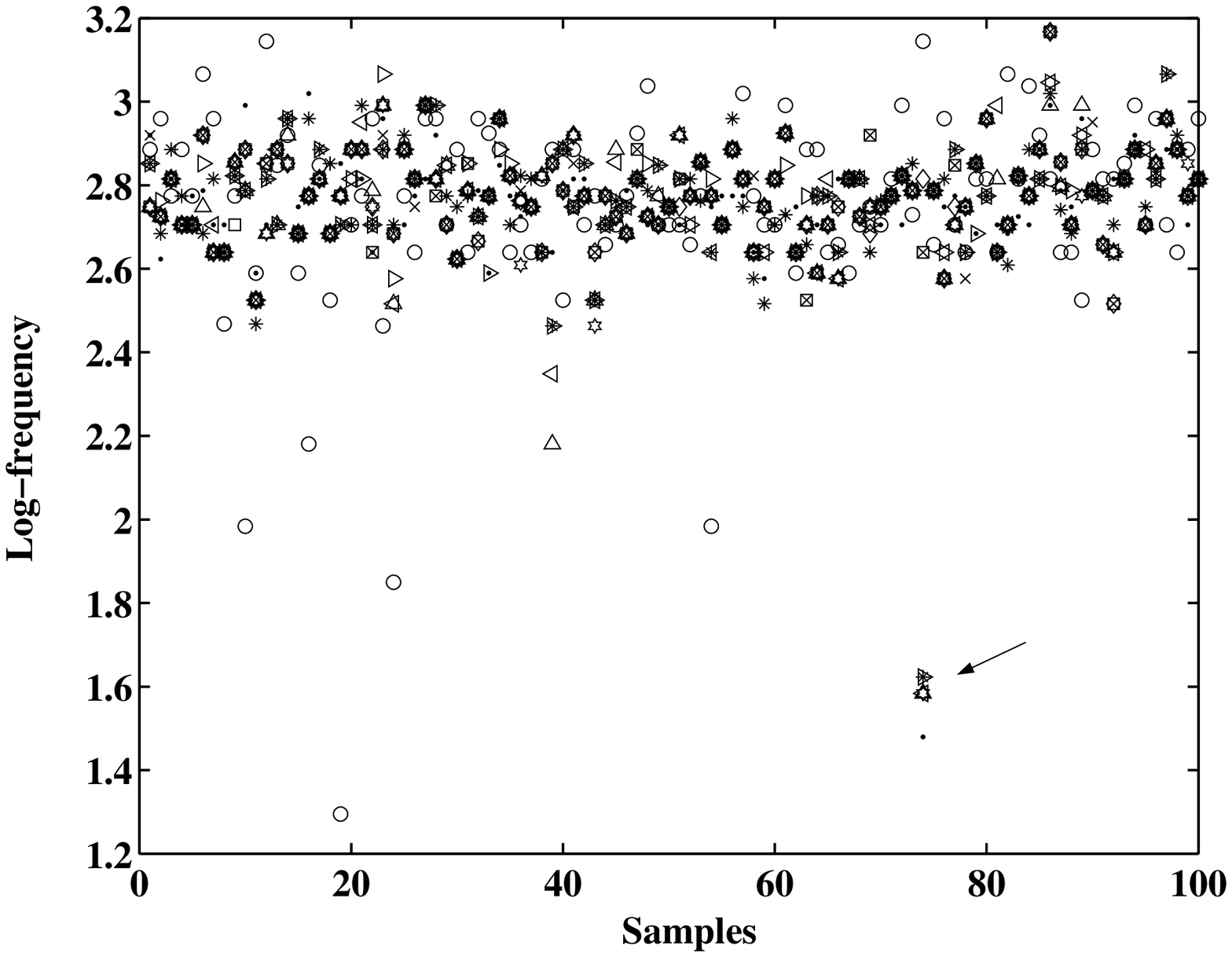,width=12cm, height=9cm}
\epsfig{file=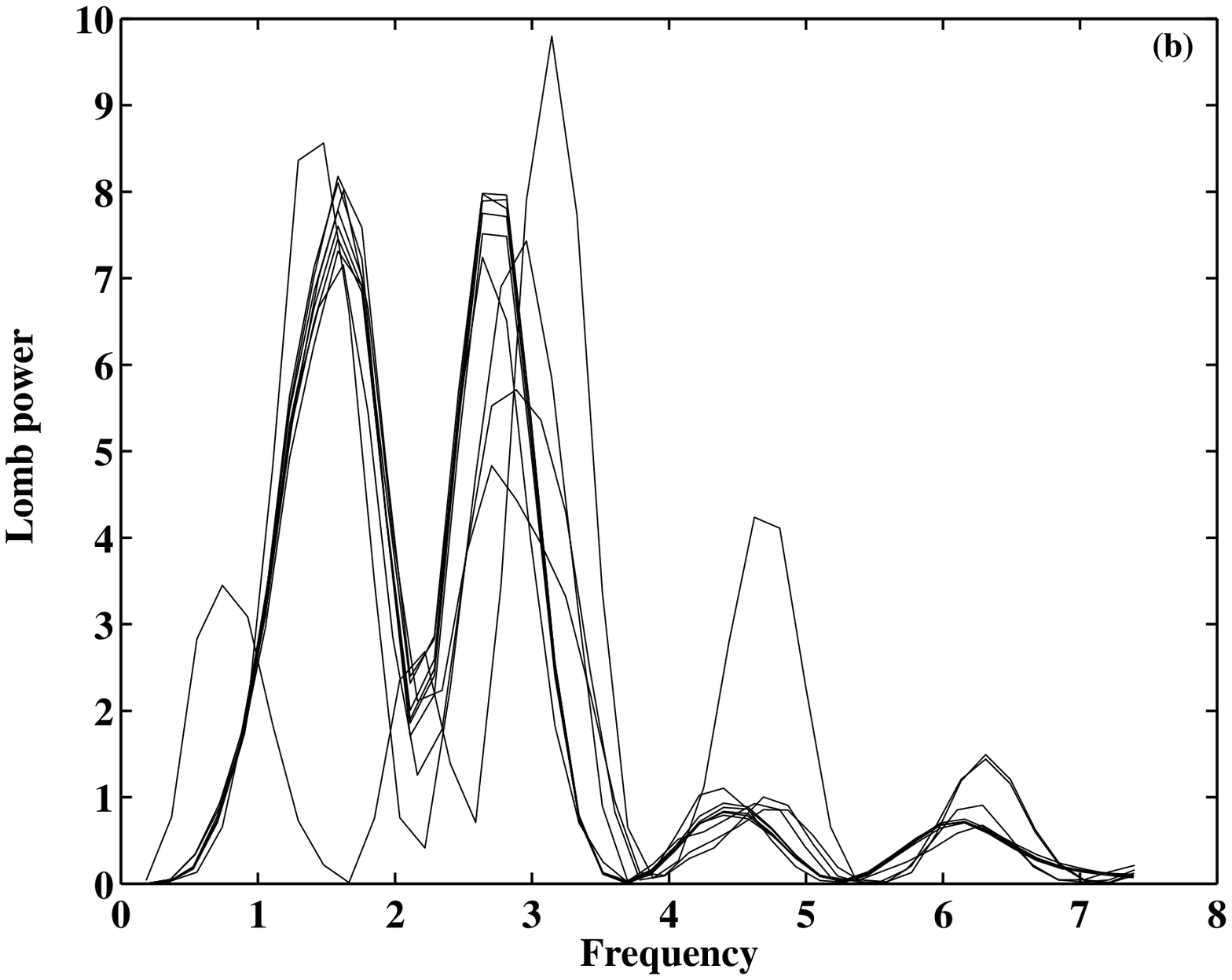,width=12cm, height=9cm}
\end{center}
\caption{(a) Leading log-frequencies $f$ of different records (numbered
on the abscissa) calculated for
different moment order $q$ marked with different symbols ($q=1$
corresponds to the open circles). The arrow indicates a large
fluctuations of the best log-frequency occurring for the 74th record.
(b) All the Lomb periodograms obtained for record 74.}
\label{Fig:fqLomb}
\end{figure}

Note also that there are pairs, say $(M=5, N_L=7)$ and $(M=6,
N_L=7)$, leading to very large standard deviations and values for
$\langle f \rangle$ that are far from any of the frequencies
$f_1$, $2 f_1, 3 f_1, 4 f_1, ...$. This apparent negative result
actually hides a remarkable confirmation of our previous
presentation. Indeed, Fig. \ref{Fig:fqLomb}(a) shows the detected
log-frequencies $f$ of the $100$ different records calculated for
different orders $q$ of the moments of the energy dissipation rate
for $(M=5, N_L=10)$. Note first that the determined frequency for
$q=1$ fluctuates more than for higher moment orders $q>1$, which
justifies our investigation of different moment orders. Secondly,
the arrow shows a large fluctuation occurring for record 74: while
most of the records are tuned to the frequency $f_2=2 f_1$, record
74 selects $f_1$ as its best log-frequency. This provides a
mechanism in which the competition between the different first
frequencies can lead to an average in between as found in Table
\ref{Tb:fMNL} for $(M=5, N_L=7)$ and $(M=6, N_L=7)$. Figure
\ref{Fig:fqLomb}(b) shows all the Lomb periodograms obtained for
record 74: we directly visualize in this case the strong
competition between the two first log-frequencies $f_1$ and $f_2$,
strengthening the case for the existence of a discrete set of
log-frequencies.

\begin{table}
\begin{center}
\begin{tabular}{|c|c|c|c|c|c|c|}
   \hline
   $\langle f
\rangle\pm\sigma_f$&$N_L=5$&$N_L=6$&$N_L=7$&$N_L=8$&$N_L=9$&$N_L=10$\\
\hline

$M=4$&$3.63\pm0.91$&$2.89\pm0.20^*$&$2.71\pm0.27^*$&$2.51\pm0.49$&$2.1
8\pm0.81$&$1.79\pm0.20^\circ$\\\hline

$M=5$&$5.22\pm0.63$&$4.32\pm0.36$&$3.76\pm0.70$&$2.99\pm0.25^*$&$2.85\
pm0.14^*$&$2.76\pm0.16^*$\\\hline

$M=6$&$5.22\pm0.64$&$4.31\pm0.36$&$3.78\pm0.69$&$3.01\pm0.28^*$&$2.85\
pm0.14^*$&$2.77\pm0.11^*$\\\hline

$M=7$&$6.15\pm0.66$&$5.72\pm0.63$&$4.82\pm0.55$&$4.37\pm0.27^\star$&$3
...90\pm0.60$&$3.13\pm0.43$\\\hline
\end{tabular}
\caption{Averaged logarithmic frequencies $\langle f \rangle$ as a
function of $M$ from $4$ to $7$ and $N_L$ from $5$ to $10$ of the
Savitsky-Golay filter.
We use $6 \times 4=24$ versions of the Savitsky-Golay filter to test for the
sensitivity of the order $M$ of the fitting polynomial and the number
$2 N_L+1$ of
points. This $24$ filters allow us to sweep broadly a large range of possible
log-frequencies and thus to access the fundamental log-frequency
$f_1$ and several of its harmonics.}
\label{Tb:fMNL}
\end{center}
\end{table}

Table~\ref{Tb:dfMNL} offers another perspective to this analysis
by showing the relative deviations $\sigma_f/\langle f \rangle$ of
the average log-frequencies shown in Table~\ref{Tb:dfMNL}. To
illustrate the robustness of the recovery of the main harmonics of
the fundamental frequency $f_1$, let us select only those
log-frequencies determined with a relative precision better than
$0.10$. We find now two classes of pairs $(M,N_L)$ of filters. The
first class contains $(4,6)$, $(4,7)$, $(5,8)$, $(5,9)$, $(5,10)$,
$(6,8)$, $(6,9)$ and $(6,10)$ and is marked by the astroid ($*$)
and corresponds to $\langle f \rangle \simeq 2.71 - 2.99$. The
second class contains (2) $(5,6)$, $(6,6)$ and $(7,8)$ and is
marked by the stars ($\star$) and corresponds to $\langle f
\rangle \simeq 4.31 - 4.37$.

\begin{table}
\begin{center}
\begin{tabular}{|c|c|c|c|c|c|c|}
   \hline
   $\sigma_f/\langle f
\rangle$&$N_L=5$&$N_L=6$&$N_L=7$&$N_L=8$&$N_L=9$&$N_L=10$\\\hline
   $M=4$&0.25&$0.07^*$&$0.10^*$&0.20&0.37&0.11\\\hline
   $M=5$&0.12&$0.08^\star$&0.19&$0.08^*$&$0.05^*$&$0.06^*$\\\hline
   $M=6$&0.12&$0.08^\star$&0.18&$0.09^*$&$0.05^*$&$0.04^*$\\\hline
   $M=7$&0.11&0.11&0.11&$0.06^\star$&0.15&0.14\\\hline
\end{tabular}
\caption{Relative deviations $\sigma_f/\langle f \rangle$ of the
averaged log-frequencies already shown in Table \ref{Tb:fMNL} as a
function of the parameters $M$ and $N_L$ of the Savitsky-Golay
filter.} \label{Tb:dfMNL}
\end{center}
\end{table}

\begin{figure}
\begin{center}
\epsfig{file=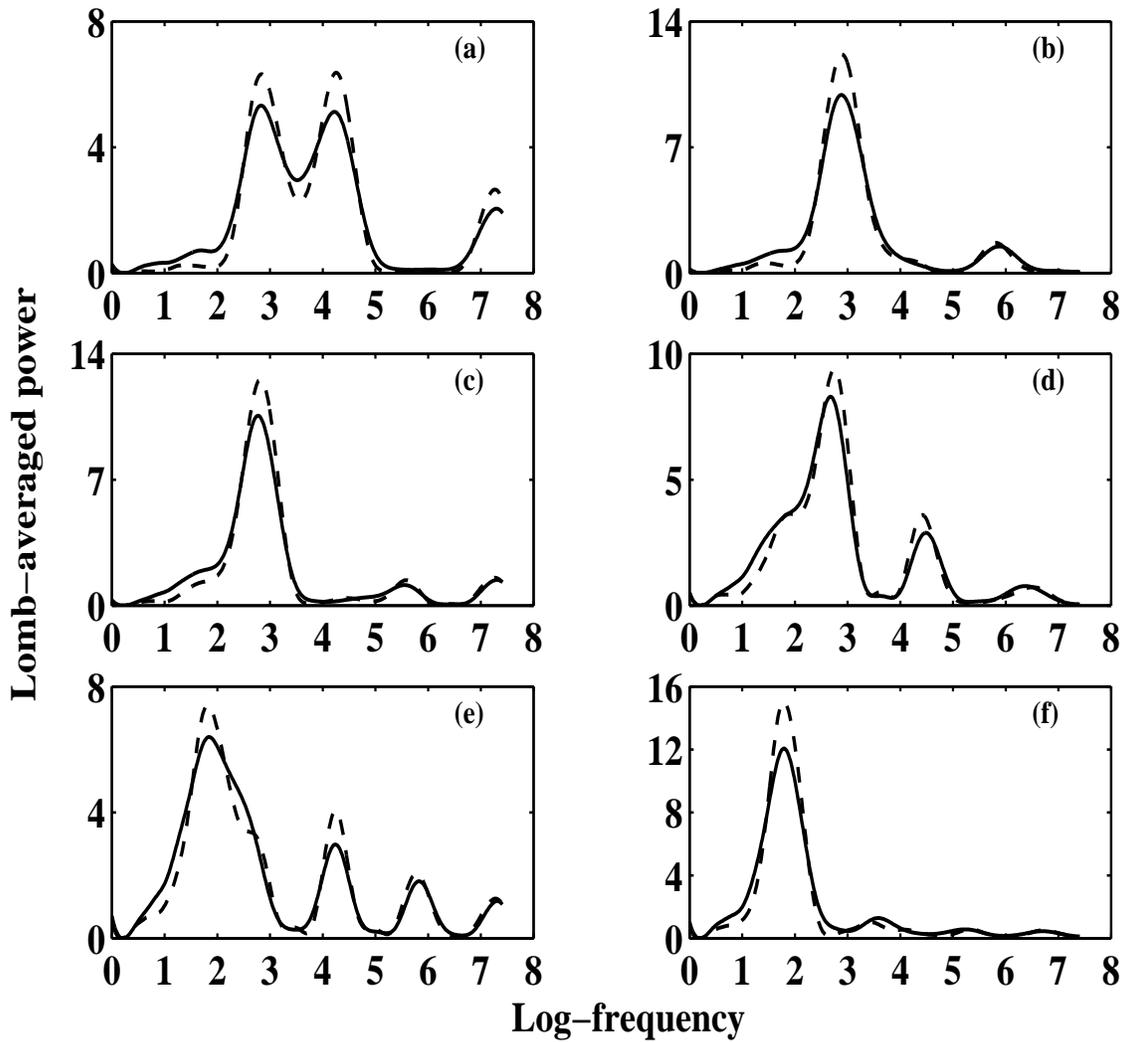,width=15cm, height=14cm}
\end{center}
\caption{Lomb periodogram obtained by averaging
over the $N=100$ (solid curve), respectively the $N=20$ (dashed line),
records of length $L=2^{17}$, respectively $L=5 \times 2^{17}$.
The order of the fitting polynomial of the Savitsky-Golay filter
is fixed at $M=4$, while $N_L$ goes from $5$ to $10$ in panels
(a) to (f).}
\label{Fig:DPLombM4NL}
\end{figure}

\begin{figure}
\begin{center}
\epsfig{file=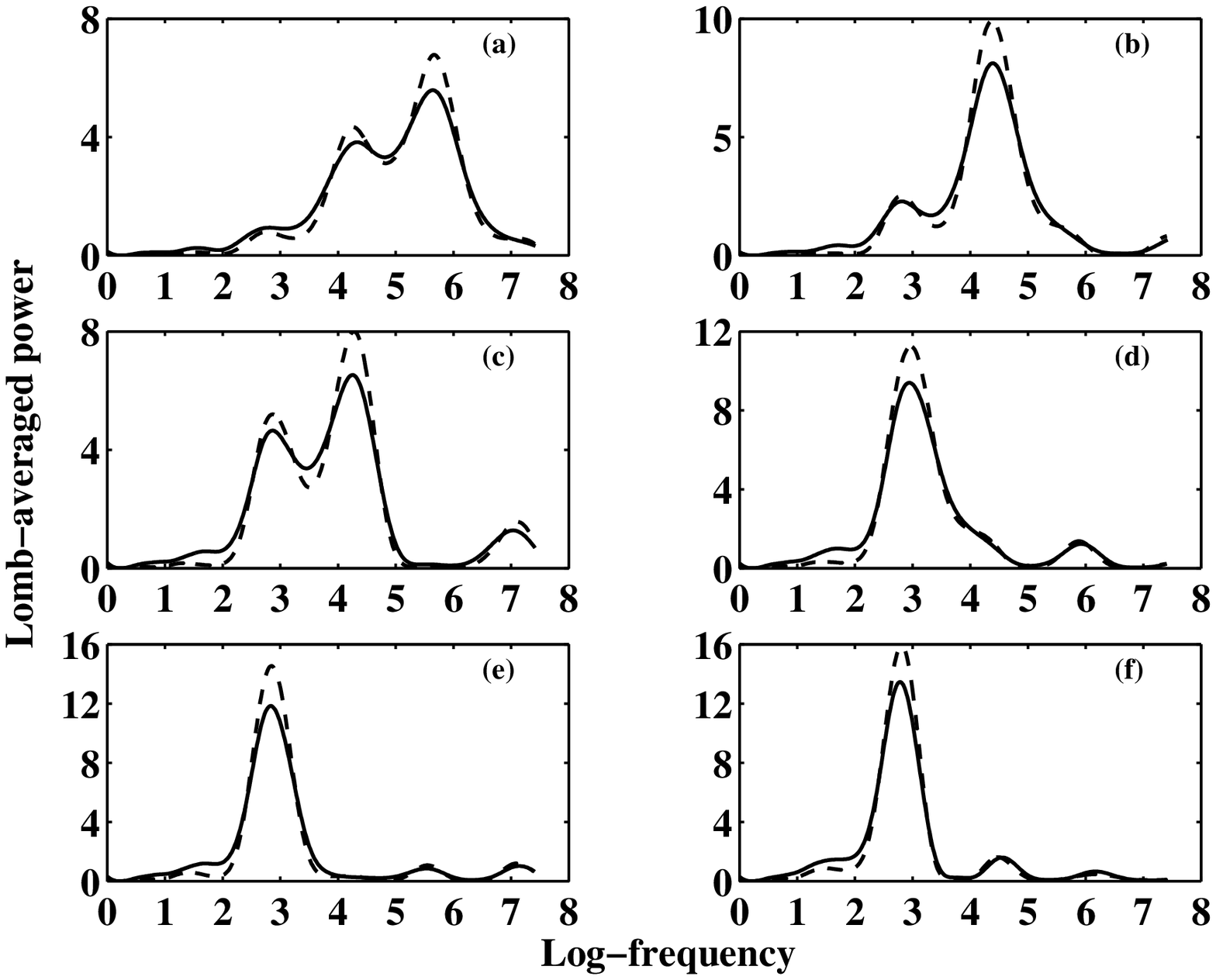,width=15cm, height=14cm}
\end{center}
\caption{Same as Fig. \ref{Fig:DPLombM4NL} with the order of the
fitting polynomial of the Savitsky-Golay filter fixed at $M=5$.}
\label{Fig:DPLombM5NL}
\end{figure}

Increasing the size $L$ of the records improves the quality of the signal.
This is first quantified by Figs.~\ref{Fig:DPLombM4NL}-\ref{Fig:DPLombM7NL}
in which the dashed lines corresponding to $L=5 \times 2^{17}$ give larger
and thinner peaks that the continuous lines obtained for $L=2^{17}$
(see below).
This confirms the role of our canonical averaging procedure which
improves in quality as the number of samples used in the averaging increases.
Table \ref{Tb:EffectL} gives the log-frequencies
$\langle f \rangle$ and their corresponding relative
deviations $\sigma_f/\langle f \rangle$ as a function of $M$ and
$N_L$ of the Savitsky-Golay filter for the $N=20$ records
of length $L=5 \times 2^{17}$. Comparing Table \ref{Tb:EffectL} ($L=5
\times 2^{17}$)
with Tables
\ref{Tb:fMNL} and \ref{Tb:dfMNL} ($L=2^{17}$), we find that a larger
$L$ leads to
a clearer determination of the harmonics log-frequencies. For
instance, keeping all frequencies such that the relative error
$\sigma_f/\langle f \rangle$
is less than $0.1$, we find that all the determined log-frequencies
now very close
the exact values of the harmonics of $f_c = 1.44$. This confirms an important
expectation of our canonical averaging procedure: using more samples
will definitely improve its efficiency for extracting some hidden
fundamental log-frequency and/or it harmonics.

\begin{table}
\begin{center}
\begin{tabular}{|c|c|c|c|c|c|c|}
   \hline
   $\langle f \rangle(\sigma_f/\langle f
\rangle)$&$N_L=5$&$N_L=6$&$N_L=7$&$N_L=8$&$N_L=9$&$N_L=10$\\\hline

$M=4$&3.68(0.20)&2.90(0.03)&2.82(0.04)&2.67(0.08)&1.92(0.18)&1.78(0.05
)\\\hline

$M=5$&5.38(0.11)&4.38(0.02)&4.07(0.12)&2.96(0.03)&2.84(0.02)&2.82(0.02
)\\\hline

$M=6$&5.53(0.08)&4.38(0.06)&4.06(0.12)&2.96(0.04)&2.84(0.03)&2.80(0.02
)\\\hline

$M=7$&6.34(0.11)&5.74(0.02)&4.72(0.10)&4.40(0.02)&4.33(0.02)&2.90(0.04
)\\\hline
\end{tabular}
\caption{Log-frequencies $\langle f \rangle$ and relative
deviations $\sigma_f/\langle f \rangle$ as a function of $M$ and
$N_L$ for the records of length $L=5 \times 2^{17}$.} \label{Tb:EffectL}
\end{center}
\end{table}

Figures~\ref{Fig:DPLombM4NL}-\ref{Fig:DPLombM7NL} summarize our
analysis by presenting the Lomb periodogram obtained by averaging
over the $N=100$ (respectively $N=20$) records of length
$L=2^{17}$ (respectively $L=5 \times 2^{17}$), for each of the
$24$ Savitsky-Golay filters described above. All the five
harmonics of the fundamental frequency $f_1$ discussed above are
seen to correspond to the main peaks of the Lomb periodograms with
a good approximation. It is expected that their amplitude should
vary with the filter, in response to the frequency selection of
each filter. It is quite remarkable however to obtain such a
strong coherence of the peaks over all the filters. This coherence
is further demonstrated by taking the average over all $24 \times N$ Lomb
periodograms ($24$
filters for each of the $N$ records) which is shown in Fig.
\ref{FigLomb2400Ave}. Notice the good agreement between the values
of the log-frequencies corresponding to the peaks and the
prediction from DSI with a root scaling ratio $\gamma = 2$ (shown
as the dashed vertical lines). We attribute the only significant
deviation observed for the first harmonics $f_1$ to a combination
of two factors: (1) the amplitude of the Lomb peak is small and
its maximum is thus mechanically biased by the existence of the
second large peak at $f_2$; (2) as we have seen, different filters
have different bandwidth and the averaging of a relatively low
signal will lead to unavoidable distortions.

\subsection{Statistical significance \label{sec:Evidence}}

The statistical significance of a Lomb peak requires in principle the
specification of the properties (distribution function and
dependence) of the noise decorating the signal. The simplest
and most often used noise is the Gaussian white noise for which the
statistical significance level of the highest Lomb peak can
be determined analytically \cite{Scarg1982, Horne1986, Press1996}. It
gives an approximately exponentially decaying false-alarm
probability:
\begin{equation}
Pr \left(  { > z} \right) = 1 -{\left( {1 - e^{ - z} } \right)}^{M'}
\sim M' e^{ - z}~, \label{EqNumRecipe1}
\end{equation}
where $M'$ is proportional to the number of data points.

\begin{figure}
\begin{center}
\epsfig{file=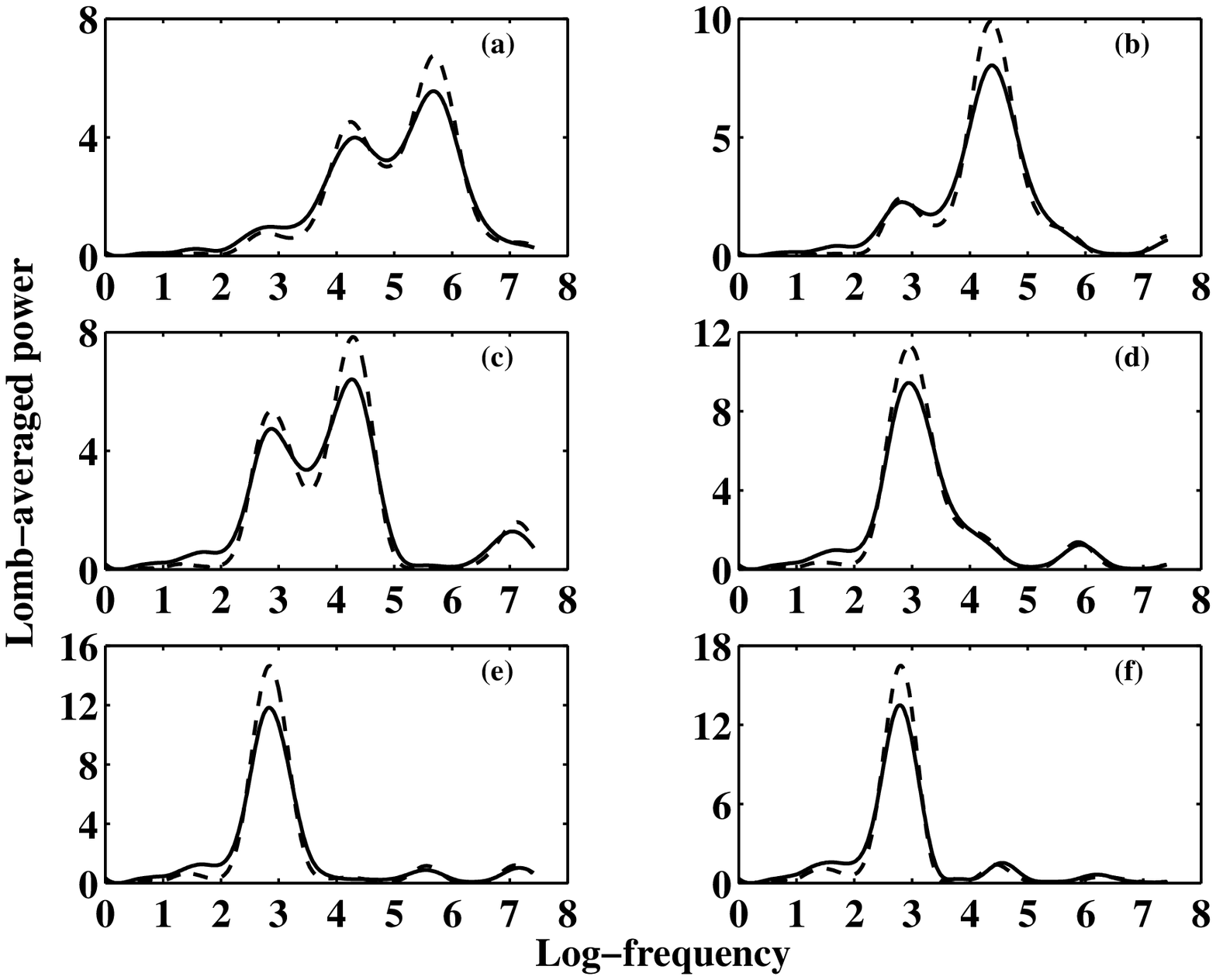,width=15cm, height=14cm}
\end{center}
\caption{Same as Fig. \ref{Fig:DPLombM4NL} with the order of the
fitting polynomial of the Savitsky-Golay filter fixed at $M=6$.}
\label{Fig:DPLombM6NL}
\end{figure}

\begin{figure}
\begin{center}
\epsfig{file=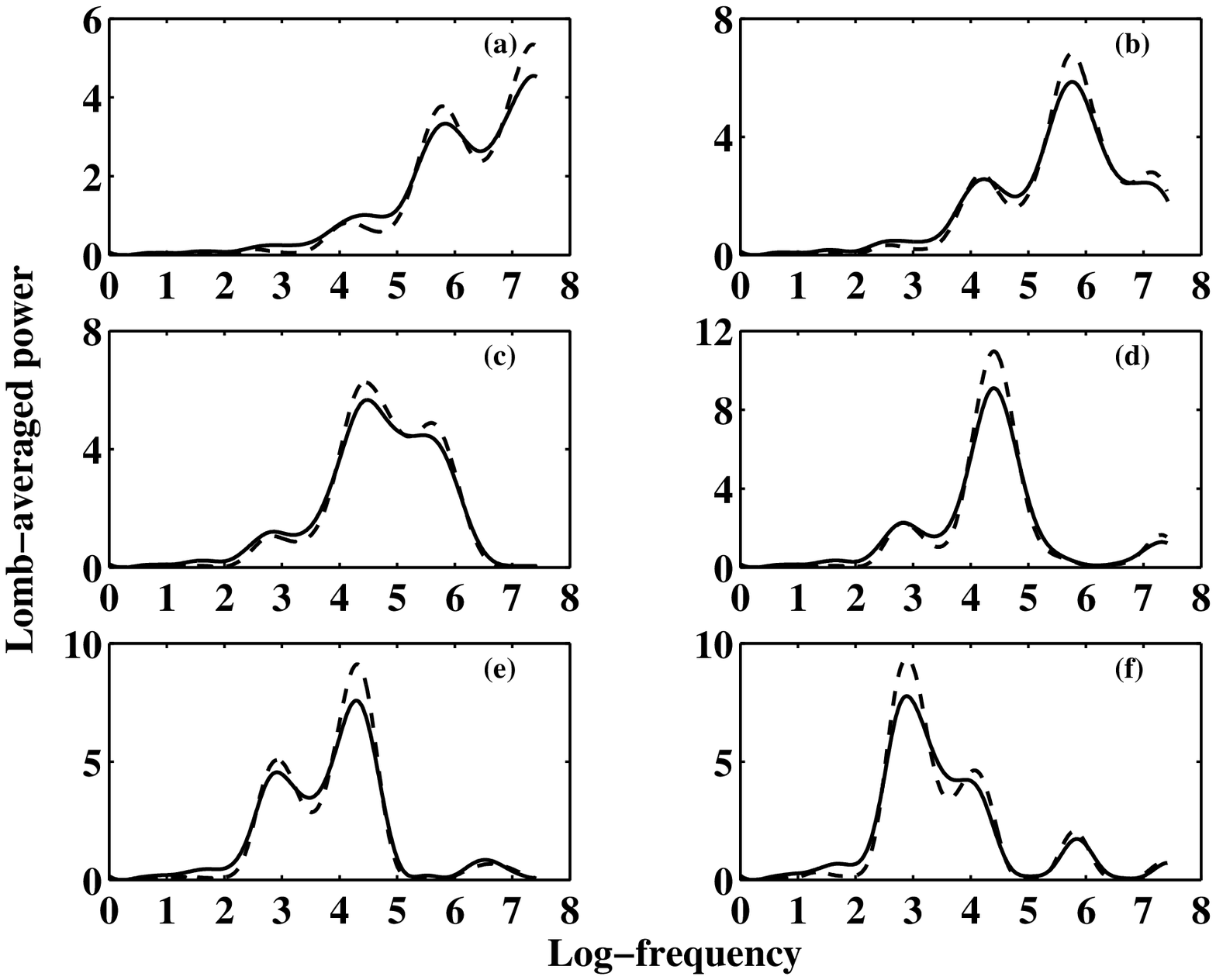,width=15cm, height=14cm}
\end{center}
\caption{Same as Fig. \ref{Fig:DPLombM4NL} with the order of the
fitting polynomial of the Savitsky-Golay filter fixed at $M=7$.}
\label{Fig:DPLombM7NL}
\end{figure}

\begin{figure}
\begin{center}
\epsfig{file=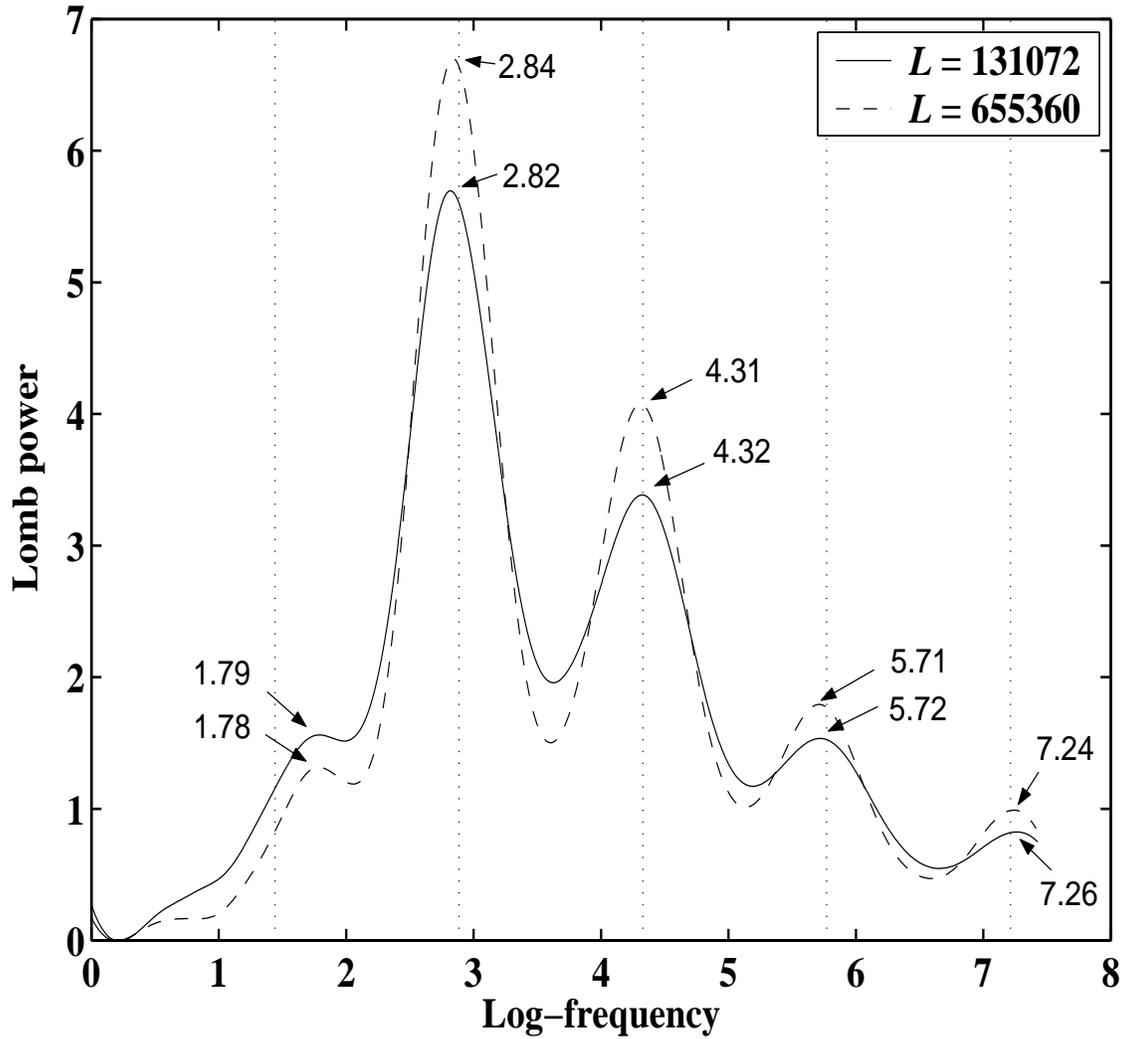,width=15cm, height=14cm}
\end{center}
\caption{Average of all $2400$ (continuous line) (resp. $480$ (dashed line))
Lomb periodograms shown in
Figs.~\ref{Fig:DPLombM4NL}-\ref{Fig:DPLombM7NL} over all $24$ filters
and all records.
The vertical dashed lines correspond to log-frequencies equal respectively to
$f_1=1.44, f_2= 2 f_1 =2.89, f_3=3 f_1=4.33$ and $f_4=4 f_1 = 5.77$.
These values correspond respectively to
increasing harmonics of a fundamental frequency $f_1 = 1/\ln \gamma$
associated with the scale ratio $\gamma = 2$.
}
\label{FigLomb2400Ave}
\end{figure}

In practice, noise is almost never Gaussian white noise with often
much fatter tails of the distribution and possible short or
long-range correlations.  Extensive numerical simulations have
investigated the impact of heavy-tailness, of correlations and of
the interplay between them on the significance level of a Lomb
periodogram peak \cite{Zhou2001b}. It was found that the worst
case leading to high false-alarm probabilities correspond to a
persistent fractional Brownian noise with large Hurst exponent.
Consider for instance the highest Lomb peak with height $16.5$
presented in Fig.~\ref{Fig:DPLombM6NL}(f) as the dashed line,
obtained with $52$ data points. According to the simulations of
Ref.~\cite{Zhou2001b}, this can be translated into a false-alarm
probability of $0.1\%$ and a statistical confidence of $99.9\%$,
ever for the worst case of a strongly persistent fractional
Gaussian noise with Hurst exponent $H=0.9$. For a Lomb peak of
height $10$ as shown in Fig.~\ref{Fig:DPLombM6NL}(b) as the dashed
lines, the confidence level is about $99.95\%$ for a fractional
Brownian noise with $H=0.7$ and $\sim 0.99$ with $H=0.8$.

\begin{figure}
\begin{center}
\epsfig{file=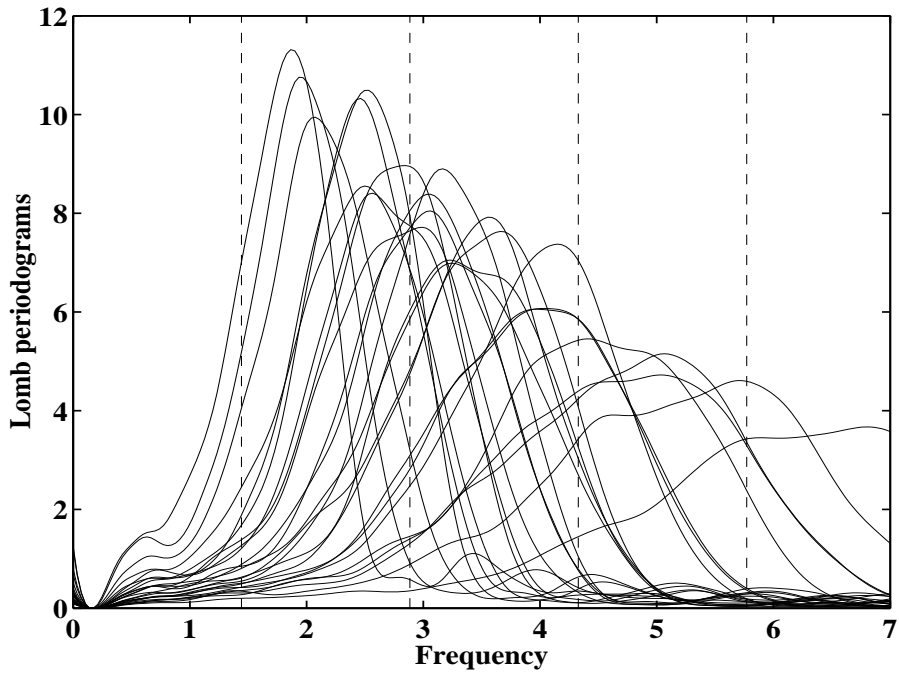,width=12cm, height=9cm}
\end{center}
\caption{Lomb periodograms averaged over $100$ samples of
fractional Gaussian noises with the
Hurst exponent $H = 0.99$, each set having 61 data points. One average
Lomb periodogram is shown for each of the $24$ values of the pair of
parameters of the
Savitsky-Golay filter:
the order $M$ of the fitting polynomial of the Savitsky-Golay filter
ranges from $4$ to $7$, while $N_L$ varies from $5$ to $10$. These
average Lomb periodograms should be compared with
Figs.~\ref{Fig:DPLombM4NL}-\ref{Fig:DPLombM7NL}.}
\label{Fig:fGnDSILomb}
\end{figure}

\begin{figure}
\begin{center}
\epsfig{file=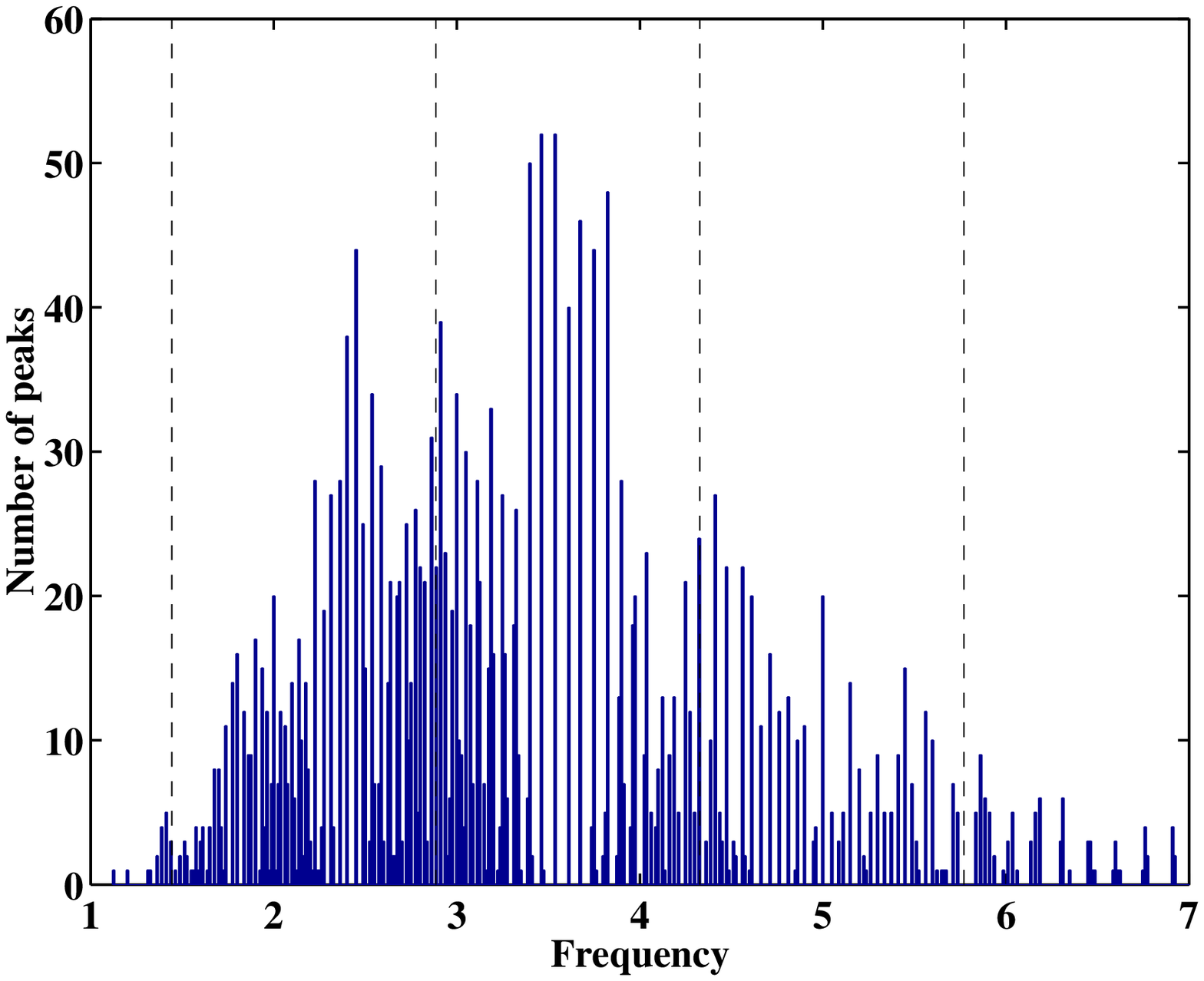,width=12cm, height=9cm}
\end{center}
\caption{Histograms of all log-frequencies extracted from
the largest peaks of the $2400$ Lomb periodograms of synthesized
fractional Gaussian noises with $H = 0.99$.}
\label{Fig:fGnDSIHist}
\end{figure}

We have performed additional simulations to test whether the five
log-frequency harmonics obtained from the Lomb peaks in
Figs.~\ref{Fig:DPLombM4NL}-\ref{Fig:DPLombM7NL}
of the turbulent data could
result from the strongest case of spurious log-periodicity
\cite{Zhou2001b,Huang2000b},
namely when the noise exhibits a strong persistence with long-range
correlations. We would like to compare the probability for obtaining
a large Lomb peak
in a purely synthetic fractional Brownian noise with a large
Hurst exponent, using exactly the canonical averaging
procedure used for analyzing the turbulence data.
It is known \cite{Zhou2001b,Huang2000b}
that, the larger $H$ is, the stronger are the spurious log-periodic
undulations.
Specifically, we use
fractional Gaussian noises of Hurst exponent $H = 0.99$ to replace
the original fluctuations decorating the leading power law of the
moments for each
sample and follow exactly the same analysis procedure as for the
turbulence data. We
generated $100$ data sets of fractional Gaussian noises with the
Hurst exponent $H = 0.99$, each set having 61 data points. The
Lomb periodograms averaged over  $100$ samples,
corresponding to different $M$ and $N_L$ of the
Savitsky-Golay filter, are shown in
Fig.~\ref{Fig:fGnDSILomb}. Compared with
Figs.~\ref{Fig:DPLombM4NL}-\ref{Fig:DPLombM7NL}, the main feature of the Lomb
periodograms in Fig.~\ref{Fig:fGnDSILomb} is that there are
no harmonics visible and
the extracted frequencies appear completely random. This is in strong contrast
with the analysis for the turbulence data which exalts the five log-periodic
frequency harmonics. Moreover, the histogram of the
total $2400$ log-frequencies corresponding to the highest peaks of
the $24 \times 100$
Lomb periodograms, shown in
Fig.~\ref{Fig:fGnDSIHist}, does not indicate that any frequency is playing
a special role. The same qualitative results are found when varying
the Hurst exponent $H$
between $0.5$ to $0.99$.

In summary, while the Savitsky-Golay filter leads to strong
spurious log-periodic components in the fractional Gaussian
noises with high Hurst exponent, their large dispersion from one
filter to the other
rules out that such a noise can be at the origin of the log-periodicity
uncovered from our analysis. In addition, since the fractional Gaussian
noises with high Hurst exponents are by far the processes leading to
the largest
spurious log-periodicity, this test also excludes all reasonable types of noise
investigated previously \cite{Zhou2001b,Huang2000b}, such as
for GARCH(1,1) noise with Student distributions with
different numbers of degrees of freedom, for power law noises, for
L\'evy stable noise
as well as of course for
independent Gaussian noise.

\section{Conclusions}
\label{sec:summary}

We have presented the results of a search for the existence of
log-periodic oscillations decorating the moments
of the energy dissipation rate in
experimental data at high Reynolds number ($\approx 2500$)
of three-dimensional fully developed turbulence. We have proposed
a simple dynamical representation of the Richardson-Kolmogorov
cartoon of a cascade that explains why
standard averaging techniques
erase by their construction the possible existence of log-periodic
corrections to scaling associated with a discrete hierarchy.
We have introduced a novel ``canonical'' averaging
methodology for extracting a possible log-periodic signal in turbulence
time series and have tested it on synthetic examples constructed
to capture the interplay between a weak log-periodic component
and rather strong multiplicative and phase noises.
The many tests presented confirm the remarkable observation of
statistically significant log-periodic correction to scaling, with
a prefered scaling ratio for length scales compatible with the value
$\gamma = 2$. A strong confirmation of this result is provided by the
identification of up to $5$ harmonics of the fundamental log-periodic
undulations, associated with up to $5$ levels of the underlying hierarchical
dynamical structure.

\bigskip
{\bf Acknowledgments:} The experimental turbulence data obtained
at ONERA Modane were kindly provided by Y. Gagne. We are grateful
to J. Delour and J.-F. Muzy for help in pre-processing these data.
This work was partially supported by NSF-DMR99-71475 and the James
S. Mc Donnell Foundation 21st century scientist award/studying
complex system.

\pagebreak

\section*{Appendix A: Study of the bandwidth selectivity of the
Savitsky-Golay filter
  \label{subsec:Selective}}

By construction, the Savitsky-Golay filter has a finite bandwidth, i.e.,
it not only filters out high frequencies but also enhances
preferentially an intermediate
frequency band whose determination depends on its two parameters, the
order $M$ of the fitting polynomial and the number $2 N_L +1$ over which this
fit is performed. Let us make this remark quantitative.

Consider a periodic function with four harmonics of the fundamental
frequency $f_c$
\begin{equation}
y(t) = \cos(2\pi f_ct) + \cos(4\pi f_ct) +\cos(6\pi f_ct) +
\cos(8\pi f_ct) + e, \label{Eq:Selective}
\end{equation}
where $f_c = 1/\ln2$, $t$ is evenly spaced in $[-1,1]$ with $61$ points
and the noise $e$ is normally distributed with zero mean and standard deviation
$\sigma_e=3$. This choice is reasonably close to the situation found
the turbulence data, in that several harmonics and noise compete
with similar amplitudes.

We generate $10$ times series of $61$ points with $t$ regularly
spaced in $\in [-1,1]$. For each of these times series, we choose
one pair $(M, N_L)$ and apply the Savitsky-Golay filter. We then
apply the Lomb periodogram technique on the resulting smoothed
time series to obtain a Lomb spectrum. We average the $10$ Lomb
spectra and identify the frequency of the largest peak which is
reported in the Table \ref{Tb:Selective}. Actually, we show the
ratios of the extracted frequencies to $f_c$. Remarkably, $17$
frequencies among the $24$ (i.e., $71\%$) are found to within less
than $10\%$ from one of the harmonics $f_c, 2 f_c, 3 f_c$. The $7$
remaining frequencies are within less than $20\%$ from one of the
harmonics $f_c, 2 f_c, 3 f_c$. The second observation is that,
indeed, a given filter $(M, N_L)$ amplifies selectively on average
one harmonics at the expanse of the other. Interestingly, when
increasing the noise amplitude $\sigma_e$, the fourth harmonic
$4f_c$ is also sampled by a few filters. Obviously, when the
amplitude of the noise is very large, spurious frequencies are
selected but come with a low statistical significance. With large
noise and some combinations of the amplitudes of the harmonics in
(\ref{Eq:Selective}), the artifactual frequency $f_m =1.5/\ln 2$
appears; its origin has been unravelled in
Ref.~\cite{Huang2000b,Huang2000a} as corresponding to the most
probable noise structure. For very large noise, we can also
identify a kind of effective non-linear coupling between the
frequencies leading to combinations of the harmonics
\cite{Berge1984}.
\begin{table}
\begin{center}
\begin{tabular}{|c|c|c|c|c|c|c|}
   \hline
   $\langle f
\rangle/f_c$&$N_L=5$&$N_L=6$&$N_L=7$&$N_L=8$&$N_L=9$&$N_L=10$\\\hline
   $M=4$&1.87&1.82&0.81&1.02&1.00&1.04\\\hline
   $M=5$&3.04&1.87&2.04&1.82&1.07&0.98\\\hline
   $M=6$&3.04&2.04&2.00&2.00&1.87&1.42\\\hline
   $M=7$&3.11&3.02&1.99&1.95&2.04&1.99\\\hline
\end{tabular}
\caption{Ratios of
the frequencies (divided by $f_c$) extracted from the largest peak of
the average
of $10$ Lomb periodograms of a synthetic signal with $4$ harmonics and a large
centered Gaussian white noise with standard deviation $\sigma_e =3$ as a
function of the two parameters $M$ and $N_L$ of the Savitsky-Golay filter.}
\label{Tb:Selective}
\end{center}
\end{table}

We have performed many tests with pure noise, with a single cosine plus noise
and with several harmonics plus noise. In the first case, the
frequencies obtained as the maxima of the Lomb spectra
do not exhibit any prefered value and fluctuate broadly from
realization to realization. With a single oscillatory signal, we see
a clear peak in the Lomb spectrum, broadened by the disorder. In the case
of several harmonics of a fundamental frequency, we find that one of the
harmonics in general dominates each Savitsky-Golay filter, allowing us
to detect all the relevant harmonics by using many different filters.

\pagebreak

\section*{Appendix B: A simple multifractal model with
discrete scale invariance (DSI) independent of the moment order $q$}

Our analysis of the log-periodic undulations decorating the
power law of the moments of the dissipation rate suggests that the
log-frequency of these undulations is independent of the order
of the moments used in the analysis. Here, we propose a simple
multifractal model with DSI which exhibits this property.

Consider a simple cascade process of the type introduced by the
Russian school and studied by Mandelbrot (see \cite{Frisch1978}
and references therein), with however one simple modification
emphasizing DSI. Each large eddy ${\bf{\mathcal{F}}}$ of size $r$
is assumed to transfer
the fraction $m_i$ of its energy to
to $n$ small eddies $\bf{{\mathcal{F}}_i}$ of
size $r_i=r/\gamma_i$, $i=1,2,\cdots,n$, where the pair $(m_i, \gamma_i)$
each are one of $n$ different
fixed values $(m_1, \gamma_1),(m_2, \gamma_2), ..., (m_n, \gamma_n)$
respectively.
Each small eddy of size $r_i$ in turn
transfers a fraction $m_j$ to each of its $n$ daughters of
size $r_i/\gamma_j$, where $m_j$ and $\gamma_j$ are drawn
from the same fixed set  $(m_1, \gamma_1),(m_2, \gamma_2), ..., (m_n,
\gamma_n)$.
This deterministic multiplicative cascade
process continues down to the dissipation scale and
creates a self-similar multifractal
measure with multipliers and characteristic scaling
ratios $m_i$ and $\gamma_i$, respectively. We do not restrict the
model to obey the space-filling condition
$\sum_{i=1}^n 1/\gamma_i = 1$  \cite{Frisch1978}.

Using the self-similarity of the cascade process, we can write
\begin{equation}
M_q({\bf{\mathcal{F}}};\ell) = \sum_{i=1}^n
M_q({\bf{\mathcal{F}}}_i;\ell)~, \label{Eq:MqAdd}
\end{equation}
where $M_q({\bf{\mathcal{F}}};\ell)$ is the $q$th-order moment of
the energy dissipation rate defined on eddy ${\bf{\mathcal{F}}}$.
According to the self-similarity of the multinomial measure, we
have
\begin{equation}
M_q({\bf{\mathcal{F}_i}};\ell) = m_i^q
M_q({\bf{\mathcal{F}}}_i;\gamma_i \ell)~. \label{Eq:MqSS}
\end{equation}
Combining Eqs.~(\ref{Eq:MqAdd}) and (\ref{Eq:MqSS}) and removing the
reference to a specific eddy ${\bf{\mathcal{F}}}$ gives the following
discrete scale
invariant equation \cite{Zhou2001c}
\begin{equation}
M_q(\ell) = \sum_{i=1}^n M_q(\gamma_i \ell) m_i^q~.
\label{Eq:MFDSI01}
\end{equation}
For statistically self-similar measures,
Eq.~(\ref{Eq:MFDSI01}) is modified by inserting in each term at the r.h.s. of
equation (\ref{Eq:MFDSI01}) the specific
probabilities of the realizations among the $n$ scenarios
at each step of the cascade,

Equation (\ref{Eq:MFDSI01}) generalizes the equation $M_q(\ell) =
m^q M_q(\gamma \ell)$ embodying the discrete scale invariance of
simple fractals constructed using the iteration of a deterministic
rule such as the triadic Cantor set or the Sierpinsky gasket. This
equation is discussed in Ref.~\cite{Lapidus2000} in relation with
so-called ``fractal strings''. Looking for a scaling solution
$M_q(\ell) \propto \ell^{\tau(q)}$, the exponents $\tau(q)$ are
the solutions of \be \sum_{i=1}^n \gamma_i^{\tau(q)} m_i^q = 1 \ee
and form an infinite countable set of complex numbers. For the
``lattice'' case where there is a common root $\gamma$ and there
are integers $k_i$ such that
\begin{equation}
\gamma_i = \gamma^{k_i}~, \label{Eq:LattCond}
\end{equation}
the solution of (\ref{Eq:MFDSI01})  can be explicited
analytically.

Substituting Eq.~(\ref{Eq:LattCond}) into Eq.~(\ref{Eq:MFDSI01}),
we have
\begin{equation}
M_q(\ell) = \sum_{i=1}^n M_q(\gamma^{k_i} \ell) m_i^q~,
\label{Eq:MFDSI02}
\end{equation}
whose solution is
\begin{equation}
M_q(\ell) =  C(\ell) \ell^{\tau(q)}~, \label{Eq:MFDSI03}
\end{equation}
where $C(\ell)$ is a universal function and the case of $C(\ell) =
{\mathtt{const.}}$ recovers the familiar power law given in
Eq.~(\ref{Eq:tau}). It is easy to verify that if $C(\ell)$ is a
periodic function of the variable $\ln(\ell)$ with the log-period
$\ln(\gamma)$, i.e.,
\begin{equation}
C(\ell) = C(\gamma^{k_i} \ell)~, \label{Eq:Cell}
\end{equation}
then Eq.~(\ref{Eq:MFDSI03}) is the solution of
Eq.~(\ref{Eq:MFDSI01}). Without loss of generality, we can rewrite
the general solution as
\begin{equation}
M_q(\ell) =  \psi_q\left[\ln(\ell)/\ln(\gamma)\right]
\ell^{\tau(q)}~. \label{Eq:MFDSI04}
\end{equation}
This implies that $M_q(\ell)$ is log-periodic. We emphasize that
$\psi_q$ is $q$-dependent, while the log-period $\ln(\gamma)$ of
$M_q(\ell)$ is independent of $q$. Another important implication is
that, given a unique prefered scaling ratio $\gamma=\exp(1/f)$,
it is possible to have a set of characteristic scaling ratios
$\gamma_i$ satisfying the lattice condition expressed by
Eq.~(\ref{Eq:LattCond}) which describe an infinite
recursion of the multifractal measure.

\pagebreak

\pagebreak

\clearpage

\clearpage

\clearpage

\clearpage

\clearpage

\clearpage

\clearpage

\clearpage

\clearpage

\clearpage

\clearpage

\clearpage

\clearpage

\clearpage

\clearpage

\clearpage

\clearpage

\begin{thebibliography}{}

\bibitem{Anifrani1995} Anifrani, J.-C., Le Floc'h, C., Sornette, D.
and Souillard, B. (1995) {\it J.Phys.I France} {\bf 5}, pp.~
631--638.

\bibitem{Anselmet1984} Anselmet, F., Y. Gagne, E.J. Hopfinger and
R.A. Antonia, High-order velocity structure functions in turbulent
shear flows, J. Fluid Mech. 140, 63-89 (1984).

\bibitem{Arneodo1995} Arn\'eodo, A., Argoul, F., Bacry, E.,
Elezgaray, J. \& Muzy, J.-F. {\it Ondelettess, multifractales et
turbulences}, Diderot Editeur, Arts et Sciences (1995).

\bibitem{Barenblatt1995} Barenblatt, G.I. and Goldenfeld, N., Does fully
developed turbulence exist? Reynolds number independence versus
asymptotic covariance, Phys. Fluids 7, 3078-3082 (1995).

\bibitem{Benzi} Benzi, R., S. Ciliberto, R. Tripiccione, C. Baudet,
C. Massaioli and S. Succi, Extended self-similarity in turbulent flows,
Phys. Rev. E 48, R29-R32 (1993).

\bibitem{Berge1984} Berge, P., Y. Pomeau and C. Vidal, Order
Within Chaos: Towards a Deterministic Approach to Turbulence
(Hermann, Paris, 1984).

\bibitem{Binder1986} Binder K. and D.W. Heermann, Spin glasses:
Experimental facts, theoretical concepts, and open questions, Rev.
Mod. Phys. 58, 801-972 (1986). (see pp. 838)

\bibitem{Couderfil} Bonn, D., Y. Couder, P.H.J. van Dam and S. Douady,
 From small scales to large scales in three-dimensional turbulence: the effect
of diluted polymers, Phys. Rev. E 47, R28-R31 (1993).

\bibitem{Castaing1997} Castaing, B. (1997) in {\it Scale invariance
and beyond}, eds. Dubrulle, B., Graner, F. \& Sornette, D., EDP
Sciences and Springer, pp.~ 225--234.

\bibitem{Dubrulle1996} Dubrulle, B., Anomalous scaling and generic
structure function in turbulence, J. Phys. France II 6, 1825-1840 (1996).

\bibitem{QuantFin1} Feigenbaum, J.A., A statistical analysis of log-periodic
precursors to financial crashes, Quantitative Finance 1 (3), 346-360 (2001)

\bibitem{Frisch1978} Frisch U, Sulem P L, Nelkin M. A simple dynamical
model of intermittent fully developed turbulence. Journal of Fluid
Mechanics 87, 719-736 (1978).

\bibitem{Frisch1996} Frisch U, Turbulence: The Legacy of A.N.
Kolmogorov (Cambridge University, Cambridge, 1996).

\bibitem{Gagne1987} Gagne, Y. Etude exp\'erimentale de l'intermittence
et des singularit\'es dans le plan complexe en turbulence d\'evelopp\'ee,
PhD thesis. University of Grenoble (1987).

\bibitem{Halsey1986} Halsey T.C., M.H. Jensen, L.P. Kadanoff,
I. Procaccia and B.I. Shraiman, Fractal measures and their
singularities: the characterization of strange sets, Phys. Rev. A
33, 1141-1151(1986).

\bibitem{Horne1986} Horne, J.H. and S.L. Baliunas, A Prescription for
period analysis of unevenly sampled time series, Astrophysical
Journal 302, 757-763 (1986).

\bibitem{Huang1997} Huang Y., G. Ouillon, H. Saleur and D.
Sornette, Spontaneous generation of discrete scale invariance in
growth models, Physical Review E 55, 6433-6447 (1997).

\bibitem{Huang2000b} Huang, Y., A. Johansen, M.W. Lee, H. Saleur and
D. Sornette,
Artifactual Log-Periodicity in Finite-Size Data: Relevance for
Earthquake Aftershocks, J. Geophys. Res. 105, 25451-25471 (2000).

\bibitem{Huang2000a} Huang, Y., H. Saleur and D. Sornette,
Reexamination of log-periodicity observed in the seismic
precursors of the 1989 Loma Prieta earthquake, J. Geophysical
Research 105, B12, 28111-28123 (2000).

\bibitem{Johansen1996} Johansen, A., D. Sornette, H. Wakita, U.
Tsunogai, W.I. Newman and H. Saleur, Discrete scaling in
earthquake precursory phenomena : evidence in the Kobe earthquake,
Japan, J.Phys.I France 6, 1391-1402 (1996).

\bibitem{Johansen1998} Johansen, A. and D. Sornette, Evidence of
discrete scale invariance by canonical averaging, Int. J. Mod.
Phys. C 9, 433-447 (1998).

\bibitem{riskcrash} Johansen, A., D. Sornette and O. Ledoit,
Predicting Financial Crashes using discrete scale invariance,
Journal of Risk 1 (4), 5-32 (1999).

\bibitem{critrup} Johansen, A. and D. Sornette,
Critical ruptures, Eur. Phys. J. B 18, 163-181 (2000).

\bibitem{Nasdaq} Johansen, A. and D. Sornette,
The Nasdaq crash of April 2000: Yet another example of log-periodicity
  in a speculative bubble ending in a crash,
  European Physical Journal B 17, 319-328 (2000).

\bibitem{Johansen2000a} Johansen, A., H. Saleur and D. Sornette,
New Evidence of Earthquake Precursory Phenomena in the 17 Jan.
1995 Kobe Earthquake, Japan, Eur. Phys. J. B 15, 551-555 (2000).

\bibitem{Johansen2000b} Johansen, A., D. Sornette and A.E. Hansen ,
Punctuated vortex coalescence and discrete scale invariance in
two-dimensional turbulence. Physica D, 2000, 138: 302-315.

\bibitem{K62} Kolmogorov, A.N., A refinement of previous hypotheses
concerning the local structure of turbulence in a viscous incompressible fluid
at high Reynolds number, J. Fluid. Dyn. 13, 82-85 (1962).

\bibitem{Kraichnan58} Kraichnan, R.H., A theory of turbulence dynamics, in
second symposium on naval hydrodynamics, 29-44, Office of Naval Research,
Washington, DC (Ref. ACR-38, 1958).

\bibitem{Kraichnan59} Kraichnan, R.H., The structure of isotropic turbulence
at very high Reynolds numbers, J. Fluid. Mech. 5, 497-543 (1959).

\bibitem{Lapidus2000} Lapidus M.L. and M. van Frankenhuysen,
Fractal Geometry and Number Theory: Complex Dimensions of Fractal
Strings and Zeros of Zeta Function (Birkh$\ddot{a}$user, Boston,
2000).

\bibitem{lvovimproved} L'vov, V.S., Podivilov, E., Pomyalov, A., Procaccia, I.
and Vandembroucq, D., Improved shell model of turbulence,
Phys. Rev. E 58, 1811-1822 (1998).

\bibitem{Meneveau1991} Meneveau C. and K.R. Sreenivasan, The
multifractal nature of turbulent energy dissipation, Journal of
Fluid Mechanics 224, 429-484 (1991).

\bibitem{Novikov1966} Novikov, E.A. (1966) {\it Dokl. Akad. Nauk SSSR}
{\bf 168/6}, pp.~1279.

\bibitem{Novikov1990} Novikov, E.A., (1990) {\it Phys. Fluids} A {\bf
2}, pp.~814--820.

\bibitem{ParisiFrisch} Parisi, G. and U. Frisch, On the singularity
structure of fully developed turbulence, in Turbulence and predictability
in geophysical fluid dynamics, Proceed. Intern. School of Physics E. Fermi,
1983, Varenna, Italy, 84-87, eds. M. Ghil, R. Benzi and G. Parisi,
North Holland, Amsterdam (1985).

\bibitem{Pazmandi1997} P\'azm\'andi F., R.T. Scalettar and G. T.
Zim\'anyi, Revisiting the Theory of Finite Size Scaling in
Disordered Systems: $\nu$ Can Be Less than 2/d, Phys. Rev. Lett.
79, 5130 (1997).

\bibitem{Press1996} Press, W., S. Teukolsky, W. Vetterling and B.
Flannery, Numerical Recipes in FORTRAN: The Art of Scientific
Computing (Cambridge University, Cambridge, 1996).

\bibitem{Renner} Renner, C., J. Peinke, R. Friedrich, O. Chanal and B. Chabaud,
On the universality of small scale turbulence, preprint physics/0109052.

\bibitem{Richardson1922} Richardson L.F., Weather Prediction by
Numerical Process (Cambridge University Press, 1922).

\bibitem{Saleur1996a} Saleur, H., C.G. Sammis and D. Sornette, Discrete scale
invariance, complex fractal dimensions and log-periodic
corrections in earthquakes, Journal of Geophysical Research-Solid
Earth 101, 17661-17677 (1996).

\bibitem{Saleur1996b} Saleur, H. and D. Sornette, Complex exponents
and log-periodic
corrections in frustrated systems, J.Phys.I France 6, n3, 327-355
(1996)

\bibitem{Scarg1982} Scargle, J.D., Study in astronomical time series analysis.
II. Statistical aspects of spectral analysis of unevenly spaced
data, Astrophysical Journal 263, 835-853 (1982).

\bibitem{Smith1986} Smith L.A., J.D. Fournier and E.A. Spiegel,
Lacunarity and intermittency in fluid turbulence, Phys. Lett. A
114, 465-468 (1986).

\bibitem{Sornette1995}  Sornette, D. and C.G. Sammis, Complex
critical exponents from renormalization group theory of
earthquakes : Implications for earthquake predictions, J.Phys.I
France 5, 607-619 (1995)

\bibitem{Sornette1996a} Sornette, D., A. Johansen, A. Arneodo,
J.-F. Muzy and H. Saleur, Complex fractal dimensions describe the
internal hierarchical structure of DLA, Phys. Rev. Lett. 76,
251-254 (1996).

\bibitem{sorjohbouch} Sornette, D., A. Johansen and J.-P. Bouchaud,
Stock market crashes, Precursors and Replicas, J.Phys.I France 6,
167-175 (1996).

\bibitem{Sornette1998a} Sornette, D., Discrete scale invariance and
complex dimensions, Phys. Rep. 297, 239-270 (1998) (see
http://xxx.lanl.gov/abs/cond-mat/9707012 for an update to 1999).

\bibitem{Sornette1998b} Sornette, D., Discrete scale invariance in
turbulence? U. Frisch (ed.), Advances in Turbulence VII, 251-254
(Kluwer Academic Publishers, The Netherlands, 1998).

\bibitem{QuantFin2} Sornette, D. and A. Johansen,
Significance of log-periodic precursors to financial crashes,
Quantitative Finance 1 (4), 452-471 (2001)

\bibitem{Stauffer1998} Stauffer, D. and D. Sornette, Log-periodic
Oscillations for Biased Diffusion on 3D Random Lattice, Physica A
252, 271-277 (1998)

\bibitem{Tcheou1996} Tch\'eou, J.-M. \& Brachet, M.E. (1996) {\it
J.Phys.II France} {\bf 6}, pp.~ 937--943.

\bibitem{Wiseman1998b} Wiseman S. and E. Domany, Self-averaging,
distribution of pseudocritical temperatures, and finite size
scaling in critical disordered systems, Phys. Rev. E 58, 2938-2951
(1998).

\bibitem{Zhou2001b} Zhou W.-X. and D. Sornette,
Statistical Significance of periodicity and
log-periodicity with heavy-tailed correlated Noise, Preprint (2001).

\bibitem{Zhou2001c} Zhou W.-X. and D. Sornette, Discrete scale
invariance in fractals and multifractal measures, 2001, in
preparation.


\end{thebibliography}
\end{document}